\documentclass[a4paper,12pt]{article}
\title{Conformal transformations and the SLE partition function martingale}
\date{}
\author{}

\newcommand{\monh}{\mathbb{H}}
\newcommand{\monht}{\mathbb{H}_t}

\newcommand{\mony}{f}
\newcommand{\monk}{k}
\newcommand{\monq}{q}
\newcommand{\monK}{K}
\newcommand{\monu}{\mathbb{U}}

\newcommand{\tild}{\tilde}
\newcommand{\ket}[1]{| #1 \rangle}
\newcommand{\bra}[1]{\langle #1 |}
\newcommand{\aver}[1]{\langle \omega | #1 | \omega \rangle}

\newcommand{\statav}[1]{\langle #1 \rangle}
\usepackage[final]{graphicx}
\usepackage{amssymb} 
\usepackage{oldgerm} 
\begin{document}
\maketitle

\vspace{-1.2cm}

\centerline{\large Michel Bauer\footnote[1]{Email:
    bauer@spht.saclay.cea.fr} and Denis Bernard\footnote[2]{Member of
    the CNRS; email: dbernard@spht.saclay.cea.fr}}

\vspace{.3cm}

\centerline{\large Service de Physique Th\'eorique de Saclay}
\centerline{CEA/DSM/SPhT, Unit\'e de recherche associ\'ee au CNRS}
\centerline{CEA-Saclay, 91191 Gif-sur-Yvette, France}

\vspace{1.0 cm}
 
\begin{abstract}
  
  We present an implementation in conformal field theory (CFT) of
  local finite conformal transformations fixing a point. We give
  explicit constructions when the fixed point is either the origin or
  the point at infinity. Both cases involve the 
  exponentiation of a Borel subalgebra of
  the Virasoro algebra. We use this to build coherent state
  representations and to derive a close analog of Wick's theorem for the
  Virasoro algebra. This allows to compute the conformal partition
  function in non trivial geometries obtained by removal of hulls from
  the upper half plane. This is then applied to stochastic Loewner
  evolutions (SLE). We give a rigorous derivation of the equations,
  obtained previously by the authors, that connect the stochastic
  Loewner equation to the representation theory of the Virasoro
  algebra. We give a new proof that this construction enumerates all
  polynomial SLE martingales. When one of the hulls removed from the
  upper half plane is the SLE hull, we show that the partition
  function is a famous local martingale known to probabilists, thereby
  unravelling its CFT origin.

\end{abstract}

\section{Introduction}

Since its very origins, the statistical mechanics of two dimensionnal
critical systems has seen a deep interplay between physics and
mathematics. This was already true for
Onsager's solution of the 2d Ising model and the computation of the
magnetization by Yang \cite{onsayang}. 
In the 80's, the link
between physics and mathematics was mainly through representation
theory, affine Lie algebras and the Virasoro algebra playing the most
central roles. Two dimensional conformal field theories \cite{bpz}
have led to an enormous amount of exact results, including the
computation of multipoint correlators and partial
classifications. The study of multifractal properties of
conformally invariant critical clusters has been less systematic, but
has nevertheless produced a number of remarquable successes (see eg.
refs.\cite{nienhuis,cardyconf,duplan} and references therein), the
famous Cardy formula giving the probability 
for the existence of a connected cluster percolating between two
opposite sides of a rectangle in two dimensional critical percolation
\cite{cardy2} being one of the highlights.
 
\vspace{.3cm}

More recently, probability theory, stochastic processes to be precise,
have started to play an important role, due to a
beautiful connection between Brownian motion and critical clusters
discovered by Schramm \cite{schramm0}. This connection is via the
Loewner evolution equation, which
describes locally growing domains $\monK _t$ (called hulls) in the
upper half plane implicitly by prescribing the variation of the
normalized uniformizing map for the complement. In this way, the
growth of the hull is coded in a real continuous function. Taking this
function to be a Brownian sample path leads to stochastic (chordal) Loewner
evolutions (SLE) of growing hulls whose properties are those expected
for conformally invariant critical clusters. There is a single
parameter, denoted $\kappa$, which is the time scale for the Brownian
motion. This has led to important probabilistic theorems, among which
Brownian intersection exponents \cite{LSW}. Moreover, this framework
made it possible to prove in certain cases that lattice statistical
models have a conformally invariant critical behavior. For instance,
Cardy's formula is now a theorem \cite{smirnov}.
 
\vspace{.3cm}

The link between SLE and standard conformal field theory (CFT) was
obscure for several years, but recently we proposed a direct
connection \cite{bb1}. The idea is to couple CFT to SLE via boundary
conditions, namely to look at a CFT in the random geometry of the
complement of the hull in the upper half plane. The crucial
observation is that if one inserts at the origin (where the hull starts
to grow) a primary boundary operator (leading to a boundary state
$\ket{\omega}$) of appropriate weight in a CFT of appropriate central
charge, and then lets the hull grow, the corresponding conformal state
is a local martingale in the sense of probability theory, i.e a
quantity whose probabilistic average is time
independant\footnote{Under certain boundedness conditions :
  technically, \textit{nice} linear forms applied to this state are
  time independant in mean.}. In this way, many quantities computed by
probabilistic methods can be shown to be directly related to
correlation functions of CFT \cite{bb2}.
 
\vspace{.3cm}

The purpose of this paper is twofold. 
 
\vspace{.2cm}

The first is SLE independent. We give a rigorous construction of the
CFT operator implementing finite local conformal transformations
fixing a point. This amounts to show how to go from certain
subalgebras of the Virasoro algebra to a corresponding Lie group
via exponentiation.   

As a first application, we use coordinates on these groups to build
coherent state highest weight representations of the Virasoro algebra.
We observe a striking similarity with the representations of the
Virasoro algebra that appear in matrix models \cite{daDVVka}. This is a
pedestrian implementation of the geometric ideas \textit{\`a la
  Borel-Weil} presented in \cite{bb3}.

Under
some global conditions, one can multiply operators corresponding to
local conformal transformations fixing different points, leading to an
embryonic version of the Virasoro group (which is ill defined in the
CFT context~: the central extension of the group of diffeomorphisms of
the circle is not what is needed). As a byproduct, we give a theorem
which does for the Virasoro algebra what Wick's theorem does for
oscillator algebras.  This kind of computation could have been made
right at the beginning of CFT, in the 80's. It seems that certain
analogous formul\ae\ were derived at that time \cite{wieg}, but we have
not been able to trace those back in the published litterature.
 
\vspace{.2cm}

These purely algebraic considerations have applications to SLE. The
uniformization of the growing hull $\monK _t$ is given, close to the point at
infinity, by a suitably normalized local conformal transformation
$\monk _t$. This leads immediately 
to a clean definition of the conformal state $G_{\monk
  _t}\ket{\omega}$ describing the growing hull $\monK _t$. The
invertible operator $G_{\monk _t}$ is then shown to satisfy a
stochastic differential equation\footnote{In our previous papers, this
  equation was used as a heuristic definition of $G_{\monk _t}$. We
  had to leave aside analytical questions of existence of solutions,
  relying on physical intuition.} which implies that $G_{\monk
  _t}\ket{\omega}$ is a local martingale.

We give
a brief account of the proof, using the above mentioned coherent
state representations of the Virasoro algebra, that $G_{\monk
  _t}\ket{\omega}$ is the generating function of all SLE martingales
in a precise algebraic sense and that these martingales build a
certain highest weight representation of the Virasoro algebra with a
non trivial character. This is an elaboration of
\cite{bb3}.

Finally, we turn to the partition function martingale. If a CFT is
coupled via boundary conditions not only to the growing hull $\monK
_t$ but also to a fixed (deterministic) hull $A$ disjoint from $\monK
_t$, the CFT partition function contains a universal contribution
corresponding to some kind of interaction between $A$ and $\monK
_t$. This is by
definition a local martingale.  We use Wick's theorem for the Virasoro
algebra to give yet another illustration that the SLE quantities
computed by probabilists \cite{LSW} are in fact deeply rooted in CFT.
For $\kappa=8/3$, this martingale computes the probability that
$\monK_t$ never touches $A$.

\vspace{.3cm}

The previous paragraph is definitely not a claim that mathematicians
have rediscovered things that were known to theoretical physicists.
Quite the opposite is true : the discoveries of probabilists have
motivated us to go back to the foundations of CFT to realize that
maybe certain basic construction had not been given enough attention
and that some CFT jewels had been left dormant.
 
\vspace{.3cm}

{\bf Acknowledgements:}
We take this opportunity to warmly thank Wendelin Werner for many
illuminating explanations on the probabilistic and geometric intuition
motivating SLE constructions and Misha Gromov for his questions on
finite conformal transformations in conformal field theory.

 Work supported in part by EC contract number
HPRN-CT-2002-00325 of the EUCLID research training network.

\section{(Chordal) stochastic Loewner evolution}

The aim of this section is to recall basic properties of stochastic
Loewner evolutions (SLE) and its generalizations that we shall need in
the following.  Most results that we recall can be found in
\cite{schramm,lawler,LSW}. See \cite{cardynew} for a nice introduction
to SLE for physicists and \cite{wernerstflour} for pedagogical summer
school notes. 

A hull in the upper half plane $\monh =\{z \in \mathbb{C}, \Im z >
O\}$ is a bounded simply connected subset $\monK \subset \monh$
(for the usual topology of $\mathbb{C}$) such that $\monh \setminus \monK$
is open, connected and simply connected. The local growth of a family of hulls
$\monK_t$ parametrized by $t \in [0,T[$ with $\monK_0=\emptyset$ is
related to complex analysis in the following way. The complement of
$\monK_t$ in $\monh$ is a domain $\monht$ which is simply
connected by hypothesis, so that by the Riemann mapping theorem
$\monht$ is conformally equivalent to $\monh$ via a map $\mony _t$.
This map can be normalized to behave as $\mony _t
(z)=z+2t/z+O(1/z^2)$ : the $PSL_2(\mathbb{R})$ automorphism group of
$\monh$ allows to impose $\mony _t (z)=z +O(1/z)$ for large $z$, and then
the coefficient of $1/z$ is fixed to be $2t$ by a time
reparametrization. The crucial condition of \textit{local} growth leads to the
Loewner differential equation
$$
\partial_t \mony_t(z)=\frac{2}{\mony_t(z)-\xi_t}\ ,\quad \mony_{t=0}(z)=z
$$
with $\xi_t$ a real function. For fixed $z$, $\mony_t(z)$ is
well-defined up to the time $\tau_z\leq +\infty$ for which
$\mony_{\tau_z}(z)=\xi_{\tau_z}$.  Then $\monK_t=\{z\in{\monh}:\ 
\tau_z\leq t\}$.

(Chordal) stochastic Loewner evolutions is obtained \cite{schramm0} by choosing 
$\xi_t=\sqrt{\kappa}\, B_t$ with $B_t$ a normalized Brownian
motion and $\kappa$ a real positive parameter so that 
$\mathbb{E}[\xi_t\,\xi_s]=\kappa\,{\rm min}(t,s)$. Here and in the following,  
$\mathbb{E}[\cdots]$ denotes expectation value.

\section{Connection with conformal field theory}
\label{sec:concft}
The next section, which also contains basic definitions to which
the reader can refer, is devoted to a careful discussion of the
implementation of finite local conformal transformations in conformal
field theory.  In this section, we simply assume that such an
implementation is possible, and we derive a direct connection between
SLE and CFT.  

\vspace{.5cm}

SLE is defined via an
ordinary differential equation, but for our reinterpretation in terms
of conformal field theories, it is useful to define $\monk _t(z) \equiv
\mony _t(z) - \xi_t$ which satisfies the stochastic differential
equation 
$$ d \monk _t = \frac{2dt}{\monk_t}-d\xi_t.$$

We observe that the conditions at spatial infinity satisfied by
$\monk _t$ imply that its germ there, which determines it
uniquely, belongs to the group $N_-$ of germs of holomorphic
functions at $\infty$ of the form $z+\sum_{m \leq -1} f_{m} z^{m+1}$,
the group law being composition. In this way, the Loewner equations
describe trajectories on $N_-$ in a time dependent left-invariant
vector field, whose value at the identity element is 
$(2/z-\dot{\xi_t})\partial_z$. 

Due to the fact that $\xi_t$ is almost surely nowhere differentiable, this
observation has to be taken with a grain of salt. We let $f \in N_-$
act on $O_{\infty}$, the space of germs of holomorphic functions at
infinity, by composition, $\gamma_f\cdot F\equiv F \circ f$. Observe
that $\gamma_{g\circ f}= \gamma_f \cdot \gamma_g$ so this is an anti
representation. Ito's formula gives
$$d \gamma_{\monk _t}\cdot F=(\gamma_{\monk _t}\cdot
F')(\frac{2dt}{\monk_t}-d\xi_t)+
\frac{\kappa}{2}(\gamma_{\monk _t}\cdot F'')$$
from which we derive 
$$\gamma_{\monk _t}^{-1} \cdot d \gamma_{\monk _t}=dt(\frac{2}{z}\partial_z
+\frac{\kappa}{2}\partial_z^2)-d\xi_t\partial_z.$$

\vspace{.5cm}

The operators $l_n=-z^{n+1}\partial_z$ are represented in conformal
field theories by operators $L_n$ which satisfy the Virasoro algebra
$\mathfrak{vir}$
$$[L_n,L_m] = (n-m)L_{n+m} +\frac{c}{12}(n^3-n) \delta_{n+m,0} \qquad
[c,L_n]=0.$$

The representations of $\mathfrak{vir}$ are not
automatically representations of $N_-$, one of the reasons being that
the Lie algebra of $N_-$ contains infinite linear combinations of the
$l_n$'s.  However, as we shall see in the next section, highest weight
representations of $\mathfrak{vir}$ can be extended in such a way as to
become representations of $N_-$. We take this for granted for the
moment and associate to $\gamma_f$ an operator $G_f$ acting on
appropriate representations and satisfying $G_{g \circ f}=G_f G_g$ and
$$ G_{\monk _t}^{-1} d G_{\monk _t}=dt(-2L_{-2}+\frac{\kappa}{2}
L_{-1}^2)+d\xi_t L_{-1}.$$   

The basic observation is the following \cite{bb1}:

\vspace{.3cm}

Let $\ket{\omega}$ be the highest weight vector in the irreducible
highest weight representation of $\mathfrak{vir}$ of central charge
$c_{\kappa}=\frac{(6-\kappa)(8\kappa-3)}{2\kappa}$ and conformal
weight $h_{\kappa}=\frac{6-\kappa}{2\kappa}$. Then
$\mathbb{E}[G_{\monk _t}\ket{\omega}]$ is time independent.

\vspace{.3cm}

This is a direct consequence of the fact that for this special choice
of central charge and weight, the irreducible highest weight
representation is degenerate at level 2 and
$(-2L_{-2}+\frac{\kappa}{2}L_{-1}^2)\ket{\omega}=0$. Then $$dG_{\monk
  _t}\ket{\omega}=G_{\monk_t}(dt(-2L_{-2}+\frac{\kappa}{2}
L_{-1}^2)+d\xi_t L_{-1})\ket{\omega}= d\xi_tG_{\monk_t}\ket{\omega}$$
From the definition of Ito integrals, $d\xi_t$ and $G_{\monk_t}$ are
independent, so that $d\mathbb{E}[G_{\monk _t}\ket{\omega}]=0$ as
announced.

\vspace{.5cm}

This result can be interpreted as follows. Take a conformal field
theory in $\monht$. The correlation functions in this geometry can be
computed by looking at the same theory in $\monh$ modulo the
insertion of an operator representing the deformation from $\monh$ to
$\monht$. This operator is $G_{\monk_t}$. Suppose that the central charge is
$c_{\kappa}$ and the boundary conditions are such that there is a boundary
changing primary operator of weight $h_{\kappa}$ inserted at the tip
of $\monk_t$ (the existence of this tip is more or less a
consequence of the local growth condition). Then in average the correlation
functions of the conformal field theory in the fluctuating geometry
$\monht$ are time independent and equal to their value at $t=0$. 
 
\vspace{.5cm}

We call $G_{\monk _t}\ket{\omega}$ a generating function for conserved
quantities because for any time-independent bra $\bra{v}$, the scalar
$\mathbb{E}[\bra{v} G_{\monk_t}\ket{\omega}]$ is a time independent
scalar. We shall see later that in an algebraic sense, all conserved
quantities for chordal SLE are of this form. 

A word of caution is needed here. Before talking about
$\mathbb{E}[\bra{v} G_{\monk_t}\ket{\omega}]$, we should in principle
show that $\bra{v} G_{\monk_t}\ket{\omega}$ is an integrable random
variable. This is true for instance if $\bra{v}$ is a finite
excitation of $\bra{\omega}$, but this condition is far too
restrictive for probability theory and for conformal field theory as
well.

In probabilistic terms, a random variable whose Ito derivative contains
only a $d\xi_t$ contribution (no $dt$) is called a local martingale.
We shall often drop the term local, even if the notion of martingale,
though closely related to the notion of local martingale, is more
restrictive. In particular, the time independence of expectations is
always true for martingales. We refer the interested reader to the
mathematical literature \cite{karatzas}.

\section{Conformal transformations in conformal field theory}

A (rather provocative) definition of (boundary) conformal field
theory is that it is the representation theory of the Virasoro
algebra $\mathfrak{vir}$.

The Virasoro algebra has an subalgebra $\mathfrak{n}_{-}$, with
generators the $L_n$'s $n <0$, which is closely related to $N_-$, the
group of germs of conformal transformations that fix $\infty$. This is
crucial for the construction of $G_{\monk_t}$. 
Our goal in this section is to show that indeed, $N_-$ acts on
sufficiently many physically relevant representations of
$\mathfrak{vir}$ to be able to make sense of conformal field theories
in the fluctuating geometry $\monht$. 

In the same spirit, the
group $N_+$ germs of conformal transformations that fix $0$ is closely
related to the subalgebra $\mathfrak{n}_{+}$ of $\mathfrak{vir}$ with
generators the $L_n$'s $n >0$. This group will also play an important
role in the forthcoming discussion. 

\subsection{Background}

The theories we shall study will mostly be boundary
conformal field theories, and will shall talk of field or operator
without making always explicit whether the argument is in the bulk or
on the boundary.

The basic principles of conformal field theory state that the fields
can be classified according to their behavior under (local) conformal
transformations. Then the correlation functions in a region $\monu$ are
known once they are known in a region $\monu _0$ and an explicit conformal
map $f$ from $\monu$ to $\monu _0$ preserving boundary conditions is given.
Primary fields have a very simple behavior under conformal
transformations : for a bulk primary field $\varphi$ of weight
$(h,\overline{h})$,
$\varphi(z,\overline{z})dz^hd\overline{z}^{\overline{h}}$ is
invariant, and for a boundary conformal field $\psi$ of weight
$\delta$, $\psi(x)|dx|^{\delta}$ is invariant. So the statistical
averages in $\monu $ and $\monu _0$ are related by
\begin{eqnarray*}
 \statav{\cdots \varphi(z,\overline{z})\cdots \psi(x)\cdots}_{\monu} &
 = & \\
& & \hspace{-3cm} \statav{\cdots \varphi(f(z),f(\overline{z}))f'(z)^h
  \overline{f'(z)}^{\overline{h}}\cdots
  \psi(f(x))|f'(x)|^{\delta}\cdots}_{\monu_0}. 
\end{eqnarray*}

Such a behavior is described as local conformal covariance.

In a local theory, small deformations are generated by the insertion of
a local operator, the stress tensor. 
Local conformal covariance can then be rephrased 
: the stress tensor of a conformal field theory is not only conserved
and symmetric, but also traceless, so that it has only two independent
components, one of which, $T$, is holomorphic (except for singularities
when the argument of $T$ approaches the argument of other
insertions), and the other one, $\overline{T}$, is antiholomorphic (again
except for short distance singularities). The field $T$ itself is not a
primary field in general, but a projective connection~:  
$$\statav{\cdots T(z) \cdots}_\monu= \statav{\cdots
  T(f(z))f'(z)^2+\frac{c}{12}\mathrm{S}f(z)\cdots}_{\monu _0}.$$
In this formula, $c$ is the central charge and
$\mathrm{S}f(z)= \left(\frac{f''(z)}{f'(z)}\right)'-\frac{1}{2}
\left(\frac{f''(z)}{f'(z)}\right)^2$ 
is the Schwartzian derivative of $f$ at $z$.

If $\monu $ is a non empty simply connected region strictly contained in
$\mathbb{C}$, the Riemann mapping theorem states that $\monu _0$ can be
chosen to be unit disk $\mathbb{D}$ or equivalently the upper-half
plane $\mathbb{H}$ with a point a infinity added, which belongs to the
boundary. This second choice will prove most convenient for us in the
sequel. 

In boundary conformal field theory, $T$ and $\overline{T}$ are not
independent : they are related by analytic continuation. The
relationship is expressed most simply in the upper-half plane.  The
vectors fields $z^{n+1}\partial_z$ and
$\overline{z}^{n+1}\partial_{\overline{z}}$ are generators of infinitesimal
conformal transformations in $\mathbb{C}$ but only the combination
$z^{n+1}\partial_z + \overline{z}^{n+1}\partial_{\overline{z}}\equiv -\ell_n$
preserves the boundary of $\mathbb{H}$, that is, the real axis. Write
$z=x+iy$ and for a while write $T(x,y)$ for what we usually write
$T(z)$. Choosing boundary conditions such that there is no flow of
energy momentum across the boundary $x=0$, $T(x,y)$ is real along the
real axis, and by the Schwartz reflection principle has an analytic
extension to the lower half plane as $T(x,-y)\equiv
\overline{T(x,y)}=\overline{T}(x,y)$. Due to this property, most contour
integrals involving $T$ and $\overline{T}$ in the upper half plane can be
seen as contour integrals involving only $T$ but in the full complex
plane.

Using conformal field theory in $\mathbb{H}$ to express correlators in
any  simply connected region strictly contained in
$\mathbb{C}$ has another advantage : one can use the formalism of
radial quantization in a straightforward way. The statistical averages
are replaced by quantum expectation values : 
$$\statav{\cdots T(z) \cdots \varphi(z,\overline{z})\cdots
  \psi(x)\cdots}_{\mathbb{H}}=\langle \Omega | \left( \cdots
      \hat{T}(z) \cdots \hat{\varphi}(z,\overline{z})\cdots
      \hat{\psi}(x)\cdots \right)_r| \Omega \rangle .$$
In this
formula, $\ket{\Omega}$ is the vacuum and $r$
denotes radial ordering : the fields are ordered from left to right
from the farthest to the closest to the origin.  The integral $\oint
dz z^{n+1}\hat{T}(z)$ along any contour of index $1$ with respect to
$0$, defines an operator $L_n$ (note again that from the point of view
of contour integrals in the upper half plane, $L_n$ involves $T$ and
$\overline{T}$). The fact that the stress tensor is the generator of
infinitesimal conformal maps implies that
\begin{eqnarray*}
 [ L_n, \hat{\psi}(x) ] & = & \left(x^{n+1}\partial_x+\delta
 (n+1)x^n \right) \hat{\psi}(x) \\
 \left[ L_n, \hat{\varphi}(z,\overline{z}) \right]  & = & 
 \left( z^{n+1}\partial_z + h(n+1) z^n + 
 \overline{z}^{n+1}\partial_{\overline{z}}+ 
 \overline{h}(n+1)\overline{z}^n 
 \right) \hat{\varphi}(z,\overline{z}) \\
 \left[ L_n, T(z) \right] & = & \left(z^{n+1}\partial_z+ 
 2(n+1)z^n\right)T(z)
 +\frac{c}{12}(n^3-n)z^{n-2} \\
 \left[ L_n,L_m \right] & = & (n-m)L_{n+m} +\frac{c}{12}(n^3-n)
 \delta_{n+m,0}.
\end{eqnarray*}
It is no surprise that we recover the commutation relations of
$\mathfrak{vir}$. Except for the anomalous $c$-term, the commutation
relations of the $L_n$'s are those of the $\ell_n$'s. Let us take this
opportunity to recall that to preserve classical symmetries in quantum
mechanics, the crucial point is to have the symmetries act well on
operators, i.e. that the adjoint action represents the classical
symmetries. This is because the phase of states are not observables. 
Hence symmetries in quantum mechanics act projectively, and this
leaves room for central terms such as $c$ in $\mathfrak{vir}$.  

The advantage of the operatorial version of conformal field theory is
that one can use the powerful methods of representation theory,
applied to the Virasoro algebra.

\subsection{Some representation theory}

In the sequel we
denote by $\mathfrak{h}$ the (maximal)  abelian subalgebra of
$\mathfrak{vir}$ generated by $L_0$ and $c$, by $\mathfrak{n}_{-}$
(resp.  $\mathfrak{n}_{+}$) the nilpotent\footnote{Triangular would
 be more accurate, but we keep this definition by analogy with finite
 dimensional Lie algebras.} Lie subalgebra of
$\mathfrak{vir}$ generated by the $L_n$'s, $n <0$ (resp. $n >0$) and by
$\mathfrak{b}_{-}$ (resp.  $\mathfrak{b}_{+}$) the Borel Lie
subalgebra of $\mathfrak{vir}$ generated by the $L_n$'s, $n \leq 0$
(resp $n \geq 0$) and $c$.

If $\mathfrak{g}$ is any Lie algebra, we denote by
$\mathcal{U}(\mathfrak{g})$ its universal enveloping algebra. Then a
representation of $\mathfrak{g}$ is the same as a left
$\mathcal{U}(\mathfrak{g})$-module. 

Let us describe representations of $\mathfrak{vir}$ by starting with
the simplest ones, which we call positive energy representations. These
are representations whose underlying space $M$ splits as a direct sum
$M=\bigoplus_{m\geq 0} M_m$ of finite dimensional subspaces such that
$L_n$ maps $M_m$ to $M_{m-n}$ for any $m,n \in \mathbb{Z}$ (with the
convention that $M_m\equiv \{0\}$ for $m<0$) and $L_0$ is
diagonalizable on each $M_m$.

If $M$ has a positive energy, we can define the contravariant
representation of $\mathfrak{vir}$ whose underlying space is the
little graded dual of $M$, which we define as $M^*\equiv \bigoplus_{m
  \geq 0} M_m^*$, where $M_m^*$ is the standard algebraic dual of the
finite dimensional $M_m$. Observe that one can view $L_n$ acting on
$M$ as a collection of linear maps $L_n: M_m \rightarrow M_{m-n}$
indexed by $m$. For each of these maps, one can take the algebraic
transpose $^tL_n: M^*_{m-n} \rightarrow M^*_m$, defined (as usual for
finite dimensional spaces) by $\left<^tL_n y,x \right> \equiv \left<
  y,L_n x \right>$ for $(x,y)\in M_m \times M^*_{m-n}$. We define
$L_n$ acting on $M^*$ by the collection $^tL_{-n}:M^*_m \rightarrow
M^*_{m-n}$. We decide that $c$ is the same scalar on $M^*$ as on $M$.
The representation property is checked by a simple computation. Note
that $M^{**}$ is canonically isomorphic to $M$ as 
a $\mathfrak{vir}$-module.

The most important examples of positive energy representations are
highest weight modules and their contravariants.

A $\mathfrak{vir}$ highest weight module $M$ is a
representation of the Virasoro algebra which contains a vector
$v$ such that (i) $\mathbb{C}v$ is a
1-dimensional representation of $\mathfrak{h}$ and is annihilated by
$\mathfrak{n}_{+}$ and (ii) the smallest subrepresentation of $M$
containing $v$ is $M$ itself, i.e. all states in $M$ can
be obtained by linear combinations of strings of generators of
$\mathfrak{vir}$ acting on $v$. Because $\mathbb{C}v$ is a
one dimensional representation of $\mathfrak{b}_{+}$, all states in
$M$
can be obtained by linear combinations of strings of generators of
$\mathfrak{n}_{-}$ acting on $v$. On such a
representation, the generator $c$ acts on $M$ as multiplication by a
scalar, which we denote by $c$ again and call the central charge. The
number $h$ such 
that $L_0 v=hv$ is called the conformal
weight of the representation. One can write $M=\bigoplus_{m \geq 0}
M_m$ where $L_0$ acts on $M_m$ by multiplication by $h+m$,
$M_0=\mathbb{C}v$ and $M_m$ is finite dimensional with
dimension at most $p(m)$, the number of partitions of $m$. For
convenience, we define $M_m\equiv \{0\}$ for $m<0$. Then $L_n$ maps $M_m$ to
$M_{m-n}$ for any $m,n \in \mathbb{Z}$.  By construction, highest weight
cyclic modules have positive energy. 

The existence of highest weight modules for given $c$ and $h$
is ensured by a universal construction using induced representation.
Let $R(c,h)$ denote the one dimensional representation of
$\mathfrak{h}$, of central charge $c$ and conformal weight $h$. View
$R(c,h)$ as a representation of $\mathfrak{b}_{+}$ where 
$\mathfrak{n}_{+}$ act trivially. This turns  $R(c,h)$ into a
left $\mathcal{U}(\mathfrak{b}_{+})$-module. For any $\mathfrak{g}$,
$\mathcal{U}(\mathfrak{g})$ acts on itself on the left and on the
right, so by restriction, we can view $\mathcal{U}(\mathfrak{vir})$ as
a left $\mathcal{U}(\mathfrak{vir})$-module and as a right
$\mathcal{U}(\mathfrak{b}_{+})$-module. Then $V(c,h)\equiv
\mathcal{U}(\mathfrak{vir})\bigotimes_{\mathcal{U}(\mathfrak{b}_{+})}
R(c,h)$ is a left $\mathcal{U}(\mathfrak{vir})$-module, called the
Verma module with parameters $(c,h)$. As a
$\mathcal{U}(\mathfrak{n}_{-})$-module, $V(c,h)$ is isomorphic to
$\mathcal{U}(\mathfrak{n}_{-})$ itself, so the number of states in
$V(c,h)_n$ is exactly $p(n)$. Any highest weight cyclic
module $M$ with parameters $(c,h)$ is a quotient of $V(c,h)$.

The contravariant $M^*$ of a highest weight module is not always
highest weight : $\mathcal{U}(\mathfrak{vir})M^*_0$ is always
irreducible, hence is a proper
submodule of $M^*$ if $M$ is not irreducible.

\subsection{Completions}

In the following, we shall often need to deal with infinite linear
combinations of Virasoro generators. For instance, formally
$T(z)=\sum_n L_n z^{-n-2}$. So we make some new definitions.

We denote by $\overline{\mathfrak{n}}_{+}$
the formal completion of $\mathfrak{n}_{+}$
which is made of arbitrary (not necessarily finite) linear
combinations of $L_n$'s, $n >0$. The Lie algebra
structure on $\mathfrak{n}_{+}$ extends to
a Lie algebra structure on $\overline{\mathfrak{n}}_{+}$ if we define
\begin{eqnarray*}
\left[ \sum_{m > 0}a_m L_m, \sum_{n > 0}b_n L_n\right] \equiv \sum_{k >0}
\Big(\sum_{m >0, n>0, \atop m+n=k} (m-n)a_mb_n\Big)L_k.
\end{eqnarray*}
As usual with formal power series, this works because for fixed $k$,
the sum $\sum_{m >0, n>0,\atop  m+n=k}$ is a finite sum. 

We can go one step further and define $\overline{\mathfrak{vir}}_+$ as
the direct sum $\overline{\mathfrak{n}}_{+}\oplus \mathfrak{b}_{-}$,
which is still a Lie algebra with the obvious definition.

One can make analogous definitions for $\overline{\mathfrak{n}}_{-}$,
$\overline{\mathfrak{b}}_{+}$, $\overline{\mathfrak{b}}_{-}$,
$\overline{\mathfrak{n}}_{-}\oplus \mathfrak{b}_{+}$. 
 
All these Lie algebras are contained
in $\overline{\mathfrak{n}}_{-}\oplus \mathfrak{h} \oplus
\overline{\mathfrak{n}}_{+}$, but we shall not (!) try to put a Lie algebra
structure on that space. 

Note that $\mathfrak{vir}$,
$\mathfrak{n}_{-}$, $\mathfrak{n}_{+}$, $\mathfrak{b}_{-}$ and
$\mathfrak{b}_{+}$ are graded Lie algebras, so their universal
enveloping algebras are graded too (the grading should not be
confused with the filtration which exists for any Lie algebra).  We
denote by $\mathcal{U}(\mathfrak{vir})_n$,
$\mathcal{U}(\mathfrak{n}_{-})_n$, $\mathcal{U}(\mathfrak{n}_{+})_n$,
$\mathcal{U}(\mathfrak{b}_{-})_n$ and
$\mathcal{U}(\mathfrak{b}_{+})_n$ the subspace of degree $n$ in each
of the corresponding algebras.

Using the grading, one checks that
$\overline{\mathcal{U}(\mathfrak{n}_{+})}\equiv \prod _{n >0}
\mathcal{U}(\mathfrak{n}_{+})_n$, the formal
completion\footnote{Following standard practice, if $I$ is a set and
  $E_i,i\in I$ a family of vector spaces indexed by $I$, $\prod_i E_i$
  is the set theoretic product of the $E_i$, whereas $\oplus_i E_i$ is
  the subspace of $\prod_i E_i$ consisting of families with only a
  finite number of nonzero components.} of
$\mathcal{U}(\mathfrak{n}_{+})$ has a natural associative algebra
structure which extends that of $\mathcal{U}(\mathfrak{n}_{+})$.
In the same spirit
$\overline{\mathcal{U}(\mathfrak{vir})}_+\equiv \oplus_{n \leq 0}
\mathcal{U}(\mathfrak{vir})_n\bigoplus \prod_{n > 0}
\mathcal{U}(\mathfrak{vir})_n $ has a natural associative algebra
structure which extends that of $\mathcal{U}(\mathfrak{vir})$, and is
in fact isomorphic to $\mathcal{U}(\overline{\mathfrak{vir}}_+)$.

Again, one can make analogous remarks for
$\overline{\mathcal{U}(\mathfrak{n}_{-})}\equiv \prod _{n <0}
\mathcal{U}(\mathfrak{n}_{-})_n$ and
$\overline{\mathcal{U}(\mathfrak{vir})}_-\equiv \oplus_{n \geq 0}
\mathcal{U}(\mathfrak{vir})_n\bigoplus \prod_{n < 0}
\mathcal{U}(\mathfrak{vir})_n $.

If $M$ is a finite energy representation, its formal completion
$\overline{M}=\prod _m M_m $  is still a $\mathfrak{vir}$-module,
though not a finite energy one. Any positive energy
representation $M$ of $\mathfrak{vir}$ is also a representation of 
$\overline{\mathfrak{vir}}_+=\overline{\mathfrak{n}}_{+}\oplus
\mathfrak{b}_{-}$ and a $\overline{\mathcal{U}(\mathfrak{n}_{+})}$-module, whereas $\overline{M}$ is a representation of
$\overline{\mathfrak{vir}}_-= \overline{\mathfrak{n}}_{-}\oplus
\mathfrak{b}_{+}$ and a $\overline{\mathcal{U}(\mathfrak{n}_{-})}$-module.

\section{Finite deformations in conformal field theory}

\begin{figure}[htbp]
  \begin{center}
    \includegraphics[width=0.6\textwidth]{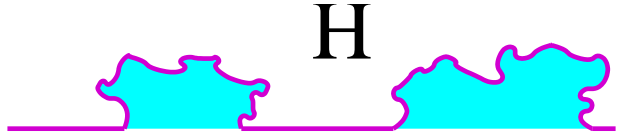}
      \caption{\em A typical hull geometry.}
      \label{fig:slebloblob}
  \end{center}
\end{figure}

Suppose now that $H$ is a domain of the type represented on
fig.(\ref{fig:slebloblob}), 
that is mapped to $\mathbb{H}$ by some conformal
transformation $f$. 

We are going to show that just as an
infinitesimal deformation is described by the insertion of an element
of the Virasoro algebra, the finite deformation that leads from the
conformal field theory on $H$ to that on 
$\mathbb{H}$ can be represented by an operator $G_f$ implementing
the map $f$:
$$\statav{\cdots \varphi(z,\overline{z}) \cdots \psi(x) \cdots }_H=\langle
  \Omega | G_f ^{-1}\left(\cdots
      \hat{\varphi}(z,\overline{z})\cdots \hat{\psi}(x) \cdots \right)_r
    G_f | \Omega \rangle .$$
This relates correlation
functions in $H$ to correlation functions in $\mathbb{H}$ where the
field arguments are taken at the same point (!)  but sandwiched inside
a conjugation by $G_f$. 

\subsection{Finite deformations around $0$}

Let $N_+$ be the space of power series
of the form $z+\sum_{m \geq 1} f_m z^{m+1}$ which have a non vanishing
radius of convergence. With words, $N_+$ is a subset of the space
$O_0$ of germs of holomorphic functions at the origin, consisting of
the germs which fix the origin and whose derivative at the origin is
$1$. In physical applications, we shall only need the
case when the coefficients are real. But in certain intermediate
constructions, it will be useful to consider the $f_m$'s as
independent commuting indeterminates (so that we forget about
convergence and deal with formal power series) : the following
statements can be translated in a straightforward way to deal with this
more general situation. 

As a set, $N_+$ is convex. Moreover, $N_+$ is a group for
composition. Our aim is to construct a group (anti)-isomorphism from $N_+$ with
composition onto  a subset $\mathcal{N}_+ \subset
\overline{\mathcal{U}(\mathfrak{n}_{+})}$ with the associative algebra
product.  The possibility to do that essentially boils down to the
fact that $\mathfrak{n}_+$ is nilpotent.
 
We let $N_+$ act on $O_0$ by $\gamma_f\cdot F\equiv F \circ f$ for $f
\in N_+$ and $F \in O_0$. This representation is faithful.  Because
$\gamma_{g\circ f}=\gamma_{f} \gamma_{g}$, we see by taking
$g=z+\varepsilon v(z)$ for small $\varepsilon$ that $\gamma_{f+\varepsilon
  v(f)}F=\gamma_{f}\cdot F +\varepsilon \gamma_{f}\cdot(v \cdot F)
+o(\varepsilon)$, where $v \cdot F(z) \equiv v(z)F'(z)$ is the
standard action of vector fields on functions. Using the Lagrange
inversion formula\footnote{With the
  convention that $\oint_0$ is an integration along a small contour of
  index $1$ around the origin, with the prefactor $(2i\pi)^{-1}$
  included, or equivalently that $\oint_0$ is taking the residue at
  the origin, a purely algebraic operation which can be performed
  without a real integration.}, we compute that for $m \geq 1$   
$$z^{m+1}=\sum_{n\geq m} f(z)^{n+1}
\oint_0 dw w^{m+1}\frac{f'(w)}{f(w)^{n+2}},$$ so that $$\frac{\partial
  \gamma_f}{\partial f_m}= \gamma_f \sum_{n\geq m} \oint_0 dw 
w^{m+1}\frac{f'(w)}{f(w)^{n+2}}\; z^{n+1}\partial_z.$$

This system of first order partial differential equations makes sense
in $\overline{\mathcal{U}(\mathfrak{n}_{+})}$ if we replace
$z^{n+1}\partial_z$ by $-L_n$. We define a connection
$$A_m \equiv  \sum_{n\geq m}
L_n \oint_0 dw  w^{m+1}\frac{f'(w)}{f(w)^{n+2}}$$
which satisfies the zero curvature condition 
\begin{equation}
  \label{eq:zerocurv}
\frac{\partial A_l}{\partial f_k}-\frac{\partial A_k}{\partial
  f_l}=[A_k,A_l].  
\end{equation}
Hence we may construct $G_f \in \overline{\mathcal{U}(\mathfrak{n}_{+})}$
for each $f \in N_+$ by solving the system
\begin{equation}
  \label{eq:pdesys}
\frac{\partial G_f}{\partial f_m}= -G_f \sum_{n\geq m} L_n \oint_0 dw
w^{m+1}\frac{f'(w)}{f(w)^{n+2}} \qquad m \geq 1.  
\end{equation}
This system is guarantied to be compatible, because the representation
of $N_+$ on $O_0$ is well defined for finite deformations $f$,
faithful and solves the analogous system. However, as the argument for
zero curvature is instructive, we give a direct proof in Appendix \ref{app:A}.

Once the compatibility conditions are checked, the existence and
unicity of $G_f$, with the initial condition $G_{f=z}$ is the
identity, is obvious : expansion of $G_{f}$ using the grading in
$\overline{\mathcal{U}(\mathfrak{n}_{+})}$ leads to a recursive system.
The group (anti)-homomorphism property is true because it is true
infinitesimally and $N_+$ is convex. 
 
As an illustration, 
$$G_f=1-f_1L_1+\frac{f_1^2}{2}(L_1^2+2L_2)-f_2L_2+\cdots$$
 Some useful general properties of $G_f$ are collected in Appendix
\ref{app:C}.

Observe that $\mathcal{N}_+$ acts by conjugation on
$\overline{\mathfrak{vir}}_+\equiv 
\overline{\mathfrak{n}}_{+}\oplus \mathfrak{b}_{-}$.  To get orientation,
let us consider the action of $f \in N_+$ not on functions but on
vector fields. First, we extend the action of $N_+$ on $O_0$ by
composition to $Q_0$, the field of fractions of $O_0$. A vector
fields $v=v(z)\partial_z$ with coefficient in $Q_0$ (i.e. a
derivations of $Q_0$) acts 
on $Q_0$ too. 

Let us consider $(\gamma_{f^{-1}}.v.\gamma_{f})F(z)$. Defining
$v_f\equiv \gamma_{f^{-1}}.v.\gamma_{f}$, a simple computation shows
that $v_fF(z)=\left(v\circ f^{-1}\right) (z) \left(f'\circ
  f^{-1}\right)(z) F'(z)$. So, as expected, $v_f$ is still a
derivation, and writing $v_f\equiv v_f(z)\partial_z$, one finds
$v_f(z)=\left(v\circ f^{-1}\right) (z) \left(f'\circ
  f^{-1}\right)(z)$. Lagrange inversion shows that
$$v_f(z)=\sum_{n\geq m} z^{n+1}\oint_0 dw w^{m+1}
\frac{f'(w)^2}{f(w)^{n+2}} \qquad \mathrm{for} \;\; v(z)=z^{m+1}.$$

Because of the correspondence
between $-z^{m+1}\partial_z$ and $L_m$, it is not surprising that, for
every $m \in \mathbb{Z}$:

\begin{eqnarray}
  \label{eq:conjl}
G_f^{-1}L_mG_f & = & \frac{c}{12} \oint_0 dw w^{m+1}Sf(w)+\sum_{n\geq m}
L_n \oint_0 dw w^{m+1} \frac{f'(w)^2}{f(w)^{n+2}} \nonumber
\\ & \equiv & L_m(f).  
\end{eqnarray}

The proof of this identity is relegated to appendix \ref{app:B}.

One can also check directly and painfully that the $L_m(f)$'s satisfy
the Virasoro algebra commutation relation with central term $c$, but
this is guarantied by the fact that $L_m(f)$ is obtained from $L_m$ by
a conjugation.

If we define a truncated stress tensor $T_l(z)\equiv \sum_{m\geq
  l}L_mz^{-m-2}$, which belongs to $\overline{\mathfrak{vir}}_+$, we have
that 
\begin{eqnarray*}
G_f^{-1}T_l(z)G_f & = & \sum_{m\geq l}L_m(f)z^{-m-2} \\ & = & \sum_{n\geq
  l}L_n\sum_{n \geq m \geq l}z^{-m-2} \oint_0 dw w^{m+1} 
 \frac{f'(w)^2}{f(w)^{n+2}}\\ & & +\frac{c}{12} \sum_{m\geq l}z^{-m-2} \oint_0
dw w^{m+1}Sf(w).  
\end{eqnarray*}
 Now let us try to let $l \rightarrow -\infty$. In
the $c$-term the $m$ summation converges to $Sf(z)$ if $z$ belongs to
the disk of convergence of $Sf(z)$. In the same way, for fixed $n$,
the $m$ summation converges to $f(z)^{-n-2}f'(z)^2$ if $z$ belongs to
the disk of convergence of $f(z)^{-n-2}f'(z)^2$. When $n$ varies, this
leads only to $2$ constraints. So, for $z$ in a non void pointed disk
centered at the origin, the infinite summations appearing for fixed
$\mathfrak{vir}$ degree in $G_f^{-1}T(z)G_f$ are absolutely convergent
and 
\begin{equation}
\label{eq:conjt}
G_f^{-1}T(z)G_f= T(f(z))f'(z)^2+\frac{c}{12}Sf(z),\end{equation} 
so we have an
operatorial version of finite deformations that has all the expected
properties. The last equation can then be extended by analytic
continuation if $f(z)$ allows it. One important lesson to draw from
this computation is that, quite naturally in fact, if the $ L_m$'s are
the basic objects and $T$ is constructed from them, changes of
coordinates act nicely only if some convergence criteria are
fulfilled. Similar consideration would apply if we would consider
the action of $G_f$ on other local fields.

Now that we have the stress tensor at our disposal, we can rewrite the
variations of $G_f$ in a familiar way : if $f$ is changed to $f+\delta
f$ with $\delta f=\varepsilon v(f)$, we find that 

$$\delta G_f =- \varepsilon G_f \oint_0 T(z)v(z) dz.$$ 

If $v$ is not just a formal
power series at the origin, but a convergent one in a neighborhood of
the origin, we can freely deform contours in this formula.

\subsubsection{Finite deformations around $\infty$}

Now, let us look at the holomorphic functions at $\infty$ instead of
$0$. So let $N_-$ be the space of power series of the form $z+\sum_{m
  \leq -1} f_{m} z^{m+1}$ which have a non vanishing radius of
convergence. We let it act on $O_{\infty}$, the space of germs of
holomorphic functions at infinity, by $\gamma_f\cdot F\equiv F \circ
f$. The adaptation of the previous computations shows that
$\frac{\partial \gamma_f}{\partial f_m}= \gamma_f \sum_{n\leq m}
\oint_{\infty} dw w^{m+1}\frac{f'(w)}{f(w)^{n+2}}\; z^{n+1}\partial_z$
where $\oint_{\infty}$ is around a small contour of index $-1$ with
respect to the
point at infinity. We transfer this relation to
$\overline{\mathcal{U}(\mathfrak{n}_{-})}$ to define an (anti)-isomorphism
from $N_-$ to $\mathcal{N}_-
\subset\overline{\mathcal{U}(\mathfrak{n}_{-})}$ mapping $f$ to $G_f$ such
that
$$\frac{\partial
  G_f}{\partial f_m}= -G_f \sum_{n\leq m} L_n \oint_{\infty} dw
w^{m+1}\frac{f'(w)}{f(w)^{n+2}}, \qquad m \leq -1.$$ 

All the previous considerations could be extended to that case.

\subsection{Dilatations and translations}

We close this section with a small extensions that, for different
reasons, demand to leave the realm of formal power series.

The first one has to do with dilatations. Up to now, we have been dealing
with deformations around $0$ and $\infty$ that did not involve
dilatation at the fixed point : $f'(0)$ or $f'(\infty)$ was unity.
Hence the operator $L_0$ appears nowhere in the above formul\ae . To
gain some flexibility in the forthcoming discussion, we decide (this
is a convention) to authorize dilatations at the origin. The operator
associated to a pure dilatation $f(z)=f'(0)z$ is $f'(0)^{-L_0}$.  One
can view a general $f$ fixing $0$ as the composition
$f(z)=f'(0)(z+\sum_m f_mz^{m+1})$ of a deformation at $0$ with
derivative $1$ at $0$ followed by a dilatation. As before, the
operators are multiplied in the opposite order, so that
$G_f=G_{f/f'(0)}f'(0)^{-L_0}$. From this formula, one checks that
eqs.(\ref{eq:conjl},\ref{eq:conjt}) remain valid even when $f$ has
$f'(0)\neq 1$. To keep the group composition
property, we demand that $f'(0)$ be real and positive.

The second extension deals with translations. Suppose that
$f(z)=f'(0)(z+\sum_m f_mz^{m+1})$ is a generic invertible germ of
holomorphic function fixing the origin ($f'(0)\neq 0$). If $a$ is in
the interior of the disk of convergence of the power series expansion
of $f$ and $f'(a)\neq 0$, we may define a new germ $f_a(z)\equiv
f(a+z)-f(a)$ with the same properties. What is the relationship
between $G_f$ and $G_{f_a}$ ? At the infinitesimal level, we compute
$\frac{df_a}{da}\vert_{a=0}=v{f}$.  The use of the Lagrange formula yields
$$v(f)=\sum_{n\geq 0}f^{n+1}\oint_0 dw \frac{f'(w)^2}{f(w)^{n+2}},$$
which implies
\begin{eqnarray*}
G_{f_a}^{-1}\frac{dG_{f_a}}{da}_{|a=0} & =  & -\sum_{n\geq 0}L_n\oint_0 dw
\frac{f'(w)^2}{f(w)^{n+2}}
\\ & = & L_{-1}f'(0)-
G_f^{-1}L_{-1}G_f.
\end{eqnarray*}
The last equality comes eq.(\ref{eq:conjl}) for $m=-1$. We conclude
that for general $a$,
$$G_{f_a}^{-1}\frac{dG_{f_a}}{da} =
L_{-1}f_a'(0)-G_{f_a}^{-1}L_{-1}G_{f_a}$$ 
This differential equation is easy to solve formally :
\begin{equation}
\label{eq:trans} 
G_{f_a}=e^{-aL_{-1}}G_fe^{f(a)L_{-1}}.\end{equation}
This formal solution has an analytic meaning at least as long as $a$
is in the interior of the disk of convergence of the power series
expansion of $f$ and $f'(a)\neq 0$ (extensions will require analytic
continuation). This is a special case of the yet to come Wick theorem
for the Virasoro algebra.

\section{An application to representation theory} \label{sec:rep}

In this section, we use the above formul\ae\ for finite deformations to
make contact with \cite{bb3}. Our goal is to construct
generalized coherent states representations of $\mathfrak{vir}$ that
will allow us to understand the structure of SLE martingales.  

\subsection{Representations associated to deformations near 0}

Suppose that $M$ is a positive energy representation of
$\mathfrak{vir}$. Then so is its dual $M^*$. Let $f$ be an element of
$N_+$.  For $(x,y)\in M \times M^*$, consider the expectation value $\left<
  G_fy,x \right>$ or $\left< G_f^{-1}y,x \right>$. From
eq.(\ref{eq:polf}) in Appendix \ref{app:C}, these expectations  are
polynomial in the coefficients of $f=z+\sum_{m \geq 1} f_m z^{m+1}$.

Take as $M$ a Verma module $V(c,h)$ and take $x\neq 0$ in the highest
weight space of $M$. Then the space $\{\left< G_fy,x \right>, y \in
M^*\}$ or $\{\left< G_f^{-1}y,x \right>, y \in M^*\}$ is the
space of all polynomials in the independent variables
$f_1,f_2,\cdots$. Indeed, choose the basis of $M$ indexed by
ordered monomials in the $L_n$'s with negative $n$, acting on the
highest weight state $x$, and the dual basis in $M^*$. Then
eq.(\ref{eq:polf}) shows that when we take for $y$ successively the
elements of the dual basis, the matrix elements $\left< G_fy,x
\right>$ or $\left< G_f^{-1} y,x \right>$ enumerate a basis of the
space of polynomials in $f_1,f_2,\cdots$. So we have two linear
isomorphisms from $M^*$ to $\monq[f_1,f_2,\cdots]$ where
$\monq$ is the preferred field of the reader ($\mathbb{Q}$ is a
minimal choice), and we can use these isomorphism to transport the
action of $\mathfrak{vir}$.

\subsubsection{The case of $G_f$} \label{sec:Gf0}

For $y \in M^*$, define $P_y\equiv \left< G_fy,x \right>$. We are
going to give formul\ae\ for  $P_{L_ny}$ as a first order differential
operator acting on $P_y$. 

The case when $n \geq 1$ is simple. 
Indeed, using formula (\ref{eq:pdesys}) for the partial derivatives
of $G_f$, one checks that 
$$ -\sum_{m \geq n} \oint_0 dz\frac{f(z)^{n+1}}{z^{m+2}}
\frac{\partial}{\partial f_m} G_f = G_f L_n$$
 So for $n \geq 1$, 
\begin{equation}
\label{eq:diffn>0f}
P_{L_ny} =  -\sum_{m \geq n}\oint_0 dz \frac{f(z)^{n+1}}{z^{m+2}}
\frac{\partial P_y}{\partial f_m}.  
\end{equation}

To deal with $n <1$, we write
$G_fL_n=(G_fL_nG_f^{-1})G_f$ and use
that $G_fL_nG_f^{-1}\in \overline{\mathfrak{n}}_{+}\oplus
\mathfrak{b}_{-}$ to decompose
$G_fL_nG_f^{-1}=(G_fL_nG_f^{-1})_{\overline{\mathfrak{n}}_{+}}
+(G_fL_nG_f^{-1})_{\mathfrak{b}_{-}}$.  

From eq.(\ref{eq:conjl}) for the compositional
inverse of $f$ , we get after a change of variable
$$
G_fL_nG_f^{-1}=-\frac{c}{12} \oint_0 dw
f(w)^{n+1}\frac{Sf(w)}{f'(w)}+\sum_{m\geq n} L_n \oint_0 dw
 \frac{f(w)^{n+1}}{w^{m+2}f'(w)} \qquad n \in \mathbb{Z}. $$
The
$\mathfrak{b}_{-}$ part contains the central charge term and the sum
$n \leq m \leq 0$. For $m <0$, $\left< L_mG_fy,x \right>=\left<
  G_fy,L_{-m}x \right>=0$ because $x$ is a highest weight state, and
$\left< L_0G_fy,x \right>=h\left< G_fy,x \right>$ because $x$ has
weight $h$.  So
\begin{eqnarray*}
\left< (G_fL_nG_f^{-1})_{\mathfrak{b}_{-}}G_fy,x\right> & = & \\ 
& & \hspace{-4cm}
\left(-\frac{c}{12} \oint_0 dw f(w)^{n+1}\frac{Sf(w)}{f'(w)}+
h\oint_0 dw f(w)^{n+1} \frac{1}{w^{2}f'(w)}\right)P_{y}.  
\end{eqnarray*}
To deal with the $\overline{\mathfrak{n}}_{+}$ part, we observe that
$G_f^{-1}(G_fL_nG_f^{-1})_{\overline{\mathfrak{n}}_{+}} G_f$ belongs
to ${\overline{\mathfrak{n}}_{+}}$ but on the other hand
$G_f^{-1}(G_fL_nG_f^{-1})_{\overline{\mathfrak{n}}_{+}}
G_f=L_n-G_f^{-1}(G_fL_nG_f^{-1})_{\mathfrak{b}_{-}}G_f$. 
Hence
$$ G_f^{-1}(G_fL_nG_f^{-1})_{\overline{\mathfrak{n}}_{+}} G_f=
-(G_f^{-1}(G_fL_nG_f^{-1})_{\mathfrak{b}_{-}}
G_f)_{\overline{\mathfrak{n}}_{+}}, \quad n<1.$$
For the second conjugation, we use eq.(\ref{eq:conjl}) for $f$ itself.
This leads to
\begin{eqnarray}\label{eq:diffn<1fd}
P_{L_ny}+\left(\frac{c}{12} \oint_0 dw
f(w)^{n+1}\frac{Sf(w)}{f'(w)}- h\oint_0 dw
 \frac{f(w)^{n+1}}{w^{2}f'(w)}\right)P_{y} & = & \\ \hspace{-2cm}
-\sum_{m =n}^0 \oint_0 dw 
\frac{f(w)^{n+1}}{w^{m+2}f'(w)}\sum_{l \geq 1} \oint_0 dz z^{m+1}
\frac{f'(z)^2}{f(z)^{l+2}} \left<  G_f L_l y,x \right>.   
\end{eqnarray}

One can express the right hand side of this formula as an explicit
differentail operator. The details are tedious and best relegated to
Appendix \ref{app:D}. The final result is that, for $n<1$,
\begin{eqnarray}\label{eq:diffn<1f}
P_{L_ny}+\left(\frac{c}{12} \oint_0 dw
f(w)^{n+1}\frac{Sf(w)}{f'(w)}- h\oint_0 dw
 \frac{f(w)^{n+1}}{w^{2}f'(w)}\right)P_{y} & = & \\ & & \hspace{-12cm}
\sum_{j \geq 1} \sum_{m =n}^0 \oint_0 dw 
\frac{f(w)^{n+1}}{w^{m+2}f'(w)}\left( f_{j-m}(j-m+1)-\sum_{k =
m}^{0}\oint_0 du \frac{u^{m+1}f'(u)^2}{f(u)^{k+2}}\oint_0 dv 
\frac{f(v)^{k+1}}{v^{j+2}}\right)\frac{\partial P_y}{\partial f_j}. \nonumber
\end{eqnarray}
Eqs.(\ref{eq:diffn>0f},\ref{eq:diffn<1f}) give the desired
representation of the action of the Virasoro algebra on $V^*(c,h)$
as first order differential operators on the space
$\monq[f_1,f_2,\cdots]$.
To be explicit, we quote the expression for a system of
generators of $\mathfrak{vir}$ :
\begin{eqnarray*}
L_2 & = & -\sum_{m \geq 2} \left(\sum_{j+k+l=m-2}f_jf_kf_l
\right)\frac{\partial}{\partial f_m}\\
L_1 & = & -\sum_{m \geq 1} \left(\sum_{j+k=m-1}f_jf_k
\right)\frac{\partial}{\partial f_m}\\
L_0 & = & h+\sum_{m \geq 1} mf_m\frac{\partial}{\partial f_m}\\
L_{-1} & = & -2f_1h+\sum_{m \geq 1}\left((m+2)f_{m+1}-2f_1(m+1)f_{m} \right)
\frac{\partial}{\partial f_m}\\
L_{-2} & = & -(f_2/2-f_1/12-f_1^2/3)c-(4f_2-7f_1^2)h \\
&  & + \mathrm{\; the \;  differential \; part}
\end{eqnarray*}
For the positive generators, the convention $f_0=1$, $f_n=0\;  n<0$ is
used within the sums. 

\subsubsection{The case of $G_f^{-1}$}

For $y \in M^*$, define $Q_y\equiv \left< G_f^{-1}y,x \right>$. We are
going to give formul\ae\ for  $Q_{L_ny}$ as a first order differential
operator acting on $Q_y$. Note that $Q_y$ is nothing but $P_y$
expressed in terms of the coefficients of the inverse (for
composition) of $f$. So in principle, the two constructions are related
by a simple change of variables. 

We use eq.(\ref{eq:conjl}) to work on $Q_{L_ny}=\left<
  (G_f^{-1}L_nG_f)G_f^{-1}y,x \right>$. Again, we write
$G_f^{-1}L_nG_f= (G_f^{-1}L_nG_f) _{\overline{\mathfrak{n}}_{+}}+
(G_f^{-1}L_nG_f)_{\mathfrak{b}_{-}}$ and use the definition of
contravariant representation on the $\mathfrak{b}_{-}$ part to keep
only the diagonal action of $\mathfrak{h}$. This leads to
\begin{eqnarray*}
Q_{L_ny} & = & \frac{c}{12} \oint_0 dw w^{n+1}Sf(w) 
+h \oint_0 dw w^{n+1}
\frac{f'(w)^2}{f(w)^{2}} \\
& & + \sum_{m\geq 1}  \oint_0 dw w^{n+1} \frac{f'(w)^2}{f(w)^{m+2}}
\left< L_m G_f^{-1}y,x \right>.  
\end{eqnarray*}
The definition of $h_m$ in eq.(\ref{eq:defhm}) and it's
characteristic property eq.(\ref{eq:prophm}) are in fact valid for
every $m \in \mathbb{Z}$. This allows to rewrite the linear
combinations of $L_m$'s as linear combinations of partial derivatives
as :
\begin{eqnarray}
Q_{L_ny} & = & \left(\frac{c}{12} \oint_0 dw w^{n+1}Sf(w) 
+h \oint_0 dw w^{n+1} \frac{f'(w)^2}{f(w)^{2}}\right)Q_{y} \\
& & \hspace{-2cm} + \sum_{m \geq \max (1,n)} \left( 
f_{m-n}(m-n+1)-\sum_{l, n \leq l \leq 0} \oint_0 du 
\frac{u^{n+1}f'(u)^2}{f(u)^{l+2}}\oint_0 dv 
\frac{f(v)^{l+1}}{v^{m+2}}\right)\frac{\partial 
Q_y}{\partial f_m} \nonumber
\end{eqnarray}

In particular
\begin{eqnarray*}
L_n & = & \sum_{m \geq 0} (m+1)f_m \frac{\partial}{\partial f_{n+m}}
  \qquad n \geq 1 \\
L_0 & = & h+\sum_{m \geq 1} mf_m\frac{\partial}{\partial f_m}\\
L_{-1} & = & 2f_1h+\sum_{m \geq 1}\left((m+2)f_{m+1}-2f_1f_{m} \right)
\frac{\partial}{\partial f_m}\\
L_{-2} & = & (f_2/2-f_1/12-f_1^2/3)c+(4f_2-f_1^2)h \\ 
 & & + \mathrm{\; the \; differential \; part}
\end{eqnarray*}

Let us note that the formula for the action of the positive generators
$L_n$, $n\geq 1$ is strikingly similar to the one that arises in
matrix models \cite{daDVVka}.

\subsubsection{Representation theoretic remarks}

By definition, a (non trivial) highest weight vector $x$ of a Verma
module $V(c,h)$ generates $V(c,h)$ when acted on by the Virasoro
generators. On the other hand, the dual $x^*$ of $x$ in $V^*(c,h)$
generates the irreducible highest weight representation of weight
$(c,h)$ when acted on by the Virasoro generators.

Hence, if $(c,h)$ is generic, i.e. if the Verma module $V(c,h)$ is
irreducible, then so is $V^*(c,h)$ and they are equivalent as
$\mathfrak{vir}$ modules. However, if $(c,h)$ is non generic, $x^*$
generates only a proper subspace of $V^*(c,h)$.

For instance, suppose that $c=\frac{(6-\kappa)(8\kappa-3)}{2\kappa}$
and $h=\frac{6-\kappa}{2\kappa}$ for some $\kappa$. Then $V(c,h)$ is
not irreducible, $(-2L_{-2}+\frac{\kappa}{2}L_{-1}^2)x$ is a singular
vector in $V(c,h)$, annihilated by the $L_n$'s, $n \geq 1$, so that it
does not couple to any descendant of $x^*$. How does this show up in
the two representations on polynomials that we constructed ? To keep
consistent notations, denote by $\mathcal{P}_n$ (resp.
$\mathcal{Q}_n$) the differential operator such that $\left< G_fL_ny,x
\right>=\mathcal{P}_n \left< G_fy,x \right>$ (resp. $\left<
  G_f^{-1}L_ny,x \right>=\mathcal{Q}_n \left< G_f^{-1}y,x \right>$)
for $y \in V^*(c,h)$. If $y$ is a descendant of $x^*$, $$\left<
  G_f^{-1}y,(-2L_{-2}+\frac{\kappa}{2}L_{-1}^2)x \right>=0.$$
On the
other hand, by copying the argument leading to the formula for
$\mathcal{P}_n$, $n\geq 1$, one checks that for $n \geq 1$ $\left<
  L_nG_f^{-1}y,x \right>=-\mathcal{P}_n\left< G_f^{-1}y,x\right>$. We
conclude that all the polynomials in $f_1,f_2,\cdots$ obtained by
acting repeatedly on the polynomial $1$ with the $\mathcal{Q}_m$'s
(they build the irreducible representation with highest weight
$(c,h)$) are annihilated by
$2\mathcal{P}_{2}+\frac{\kappa}{2}\mathcal{P}_{1}^2$. For generic
$\kappa$ there is no other singular vector in $V(c,h)$, and this leads
to a satisfactory description of the irreducible representation of
highest weight $(h,c)$ : the representation space is given by the
kernel of an explicit differential operator acting on
$\monq[f_1,f_2,\cdots]$, and the states are build by repeated
action of explicit differential operators on the highest weight state
$1$. The same argument would apply to general singular vectors.

\subsection{Representations associated to deformations near $\infty$}

The presentation parallels quite closely the case of deformations
around $0$ so we shall not give all the details. All arguments can be
adapted straightforwardly.

Again, $M$ and its dual $M^*$ are supposed to be  positive energy
representation of $\mathfrak{vir}$. But now  we take $f$ in
$N_-$.  For $(x,y)\in M \times M^*$, consider the expectation value $\left<
 y, G_fx \right>$ or $\left<y, G_f^{-1} x \right>$ \footnote{Though
 neither $G_fx$ nor $G_f^{-1} x$ is a finite excitation of $x$ in general, the
 matrix elements $\left<
 y, G_fx \right>$ and $\left<y, G_f^{-1} x \right>$ are well defined
because $y \in M^*$ is by definition a finite excitation.}. As for the
deformations around $0$, these expectations are polynomial in the
coefficients of $f=z+\sum_{m \leq -1} f_m z^{m+1}$.

As $M$, take a Verma module $V(c,h)$ and take $x\neq 0$ in the highest
weight space of $M$. The space $\{\left< y,G_fx \right>, y \in
M^*\}$ or $\{\left< y,G_f^{-1}x \right>, y \in M^*\}$ is the
space of all polynomials in the independent variables
$f_{-1},f_{-2},\cdots$.  So we have two linear
isomorphisms from $M^*$ to $\monq[f_{-1},f_{-2},\cdots]$ and 
we can use these isomorphism to transport the
action of $\mathfrak{vir}$.

\subsubsection{The case of $G_f$}

For $y \in M^*$, define $R_y\equiv \left< y, G_fx \right>$. We 
give formul\ae\ for $R_{L_ny}$ as a first order differential
operator acting on $R_y$. 
We write $\left< L_n y, G_fx \right> =
\left<  y, L_{-n} G_fx \right>$ and conjugate to obtain
$$R_{L_ny}=\frac{c}{12}\oint_{\infty} dz z^{1-n}Sf(z) R_y +\sum_{m \leq -n}
\oint_{\infty} dz z^{1-n} \frac{f'(z)^2}{f(z)^{m+2}}\left< y, G_fL_m
x \right>,$$ where $\oint_{\infty}$ is around a small contour of index $-1$ with
respect to the point at infinity.
Using the highest weight property of $x$ we get 
\begin{eqnarray*}
R_{L_ny}-\left(\frac{c}{12}\oint_{\infty} dz
 z^{1-n}Sf(z)+h\oint_{\infty} dz z^{1-n} \frac{f'(z)^2}{f(z)^{2}}
  \right) R_y & = & \\ & & \hspace{-6cm} \sum_{m \leq -1} 
\oint_{\infty} dz z^{1-n} \frac{f'(z)^2}{f(z)^{m+2}}
\left< y, G_fL_m x \right>.
\end{eqnarray*}
As in the previous sections, we may express the right hand side as an
explicit differential operator. Define, for $n \in \mathbb{Z}$, 
$$ i_n(z) \equiv z^{1-n}f'(z)-\sum_{m, \; n\leq m\leq 0}
f(z)^{1-m}\oint_{\infty} du \frac{u^{1-n}f'(u)^2}{f(u)^{2-m}},$$ which
has the property that $i_n(z)=O(1)$ and 
$$\oint_{\infty} dz z^{1-n}\frac{f'(z)^2}{f(z)^{m+2}}=\oint_{\infty} dz 
\frac{h_n(z)f'(z)}{f(z)^{m+2}} \quad \mathrm{for} \quad m=-1,-2,\cdots.$$
The $z$ expansion reads
$$i_n(z)=\sum_{m \leq -1} z^{m+1}\left( f_{m+n}(m+n+1)-\sum_{l,\; n
    \leq l\leq 0} \oint_{\infty} du
  \frac{u^{1-n}f'(u)^2}{f(u)^{2-l}}\oint_{\infty} dv
  \frac{f(v)^{1-l}}{v^{m+2}}\right).$$
This leads to the formula
\begin{eqnarray}
R_{L_ny} & = & \left(\frac{c}{12} \oint_{\infty} dw w^{1-n}Sf(w) 
+h \oint_{\infty} dw w^{1-n} \frac{f'(w)^2}{f(w)^{2}}\right)Q_{y} \\
& & \hspace{-2cm} - \sum_{m \leq \min (-1,-n)} \left(f_{m+n}(m+n+1)-
\sum_{l,\; n \leq l\leq 0} \oint_{\infty} du
\frac{u^{1-n}f'(u)^2}{f(u)^{2-l}}\oint_{\infty} dv
\frac{f(v)^{1-l}}{v^{m+2}} 
\right)\frac{\partial 
R_y}{\partial f_m} \nonumber
\end{eqnarray}
which yields
\begin{eqnarray*}
L_n & = & -\sum_{m \leq 0} (m+1)f_m \frac{\partial}{\partial f_{m-n}}
  \qquad n \geq 1 \\
L_0 & = & h-\sum_{m \leq -1} mf_m\frac{\partial}{\partial f_m}\\
L_{-1} & = & -2f_{-1}h-\sum_{m \leq -1}\left(mf_{m-1}\; \; 
  -\sum_{k+l=m-1}f_kf_l\; \; +\; \; 2f_{-1}f_{m}\right)
\frac{\partial}{\partial f_m}\\
L_{-2} & = & -cf_{-2}/2-h(4f_{-2}-3f_{-1}^2) \\ 
 & & -\sum_{m \leq -1} \left( (m-1)f_{m-2}\; \; -\sum_{j+k+l=m-2}f_jf_kf_l
   \; \; \;+ \right.\\
& & \hspace{1.7cm} \left. 3f_{-1}\sum_{k+l=m-1}f_kf_l\; \; + \; \;(4f_{-2}-3f_{-1}^2)f_m\right)
\frac{\partial}{\partial f_m}
\end{eqnarray*}

\subsubsection{The case of $G_f^{-1}$}

For $y \in M^*$, define $S_y\equiv \left< y, G_f^{-1}x \right>$. We 
give formul\ae\ for $S_{L_ny}$ as a first order differential
operator acting on $S_y$. We write $\left< L_n y, G_f^{-1}x \right> =
\left<  y, L_{-n} G_f^{-1}x \right>$.
The case $n \geq 1$ is easy. From
$$\sum_{m \leq -n} \oint_{\infty} dz\frac{f(z)^{1-n}}{z^{m+2}}
\frac{\partial}{\partial f_m} G_f^{-1} = L_{-n}G_f^{-1}$$ 
we infer that
$$S_{L_ny}=\sum_{m\leq -n} \oint_{\infty} dz\frac{f(z)^{1-n}}{z^{m+2}}
\frac{\partial S_y}{\partial f_m} \qquad n \geq 1.$$
In particular
\begin{eqnarray*}
S_{L_1y} & = & \frac{\partial S_y}{\partial f_{-1}} \\
S_{L_2y} & = & \sum_{m\leq -2} \oint_{\infty} dz\frac{1}{f(z)z^{m+2}}
\frac{\partial S_y}{\partial f_m}.  
\end{eqnarray*}

The study of the case $n <1$ follows closely the discussion in section
\ref{sec:Gf0}. As it plays no role in the application to SLE we
leave the computation to the reader. 

\subsubsection{Application to SLE martingales}

We assume that $c=\frac{(6-\kappa)(8\kappa-3)}{2\kappa}$
and $h=\frac{6-\kappa}{2\kappa}$ for some $\kappa$. Then $V(c,h)$ is
not irreducible, $(-2L_{-2}+\frac{\kappa}{2}L_{-1}^2)x$ is a singular
vector in $V(c,h)$, annihilated by the $L_n$'s, $n \geq 1$, so that it
does not couple to any descendant of $x^*$, the dual of $x$. The
descendants of $x^*$ in $V^*(c,h)$
generate the irreducible highest weight representation of weight
$(c,h)$. We denote by $\mathcal{R}_n$ (resp.
$\mathcal{S}_n$) the differential operator such that $\left<L_ny, G_fx
\right>=\mathcal{R}_n \left< y,G_fx \right>$ (resp. $\left<
  L_ny,G_f^{-1}x \right>=\mathcal{S}_n \left< y,G_f^{-1}x \right>$)
for $y \in V^*(c,h)$. Now for $n \geq 1$, $\left<y, G_fL_{-n}x
\right>=-\mathcal{S}_n\left<y, G_fx
\right>$. If $y$ is a descendant of $x^*$,
$$\left<y,G_f(-2L_{-2}+\frac{\kappa}{2}L_{-1}^2)x \right>=0$$ 
All the polynomials in $f_{-1},f_{-2},\cdots$ obtained by
acting repeatedly on the polynomial $1$ with the $\mathcal{R}_m$'s
(they build the irreducible representation with highest weight
$(c,h)$) are annihilated by
$2\mathcal{S}_{2}+\frac{\kappa}{2}\mathcal{S}_{1}^2$. For generic
$\kappa$ there is no other singular vector in $V(c,h)$, and this leads
to a satisfactory description of the irreducible representation of
highest weight $(h,c)$ : the representation space is given by the
kernel of an explicit differential operator acting on
$\monq[f_{-1},f_{-2},\cdots]$, and the states are build by repeated
action of explicit differential operators (the $\mathcal{R}_m$'s) on
the highest weight state $1$. 

We are now in position to rephrase the main results of $\cite{bb3}$
in the language of this paper.
If we take $f=\monk _t$, the coefficients $f_{-1},f_{-2},\cdots$ of
$f$ become random functions (for instance $f_{-1}$ is simply a
Brownian motion of covariance $\kappa$). One can show (see  \cite{bb3}
for details) that for fixed $t$ the coefficients
$f_{-1},f_{-2},\cdots$
seen as functions over the Wiener sample space are algebraically
independent. 

So the above computation can be interpreted as follows :
the space of polynomials of the coefficients of the expansion
of $\monk _t$ at $\infty$ for SLE$_\kappa$ can be endowed with a
Virasoro module structure isomorphic to $V^*(c_{\kappa},h_{\kappa})$.
Within that space, the subspace of martingales is a submodule
isomorphic to the irreducible highest weight representation of weight
$(c_{\kappa},h_{\kappa})$.

\section{``Wick's theorem'' for the Virasoro algebra}
\label{sec:wick}

\begin{figure}[htbp]
  \begin{center}
    \includegraphics[width=0.6\textwidth]{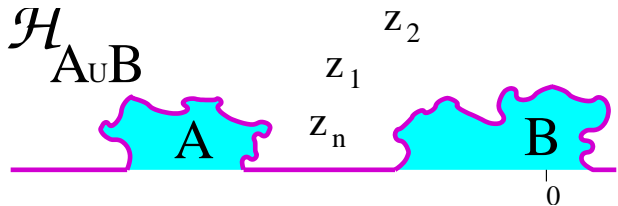}
      \caption{\em A typical two hulls geometry.}
      \label{fig:sleblob}
  \end{center}
\end{figure}

Up to now, we have only dealt with finite deformations close to $0$ or
$\infty$. These are the most natural points for radial quantization in
conformal field theory. However, this is not always convenient. 
A typical situation is as depicted in fig.(\ref{fig:sleblob}). 

We want to evaluate correlation of
operators in a geometry where the natural series at $0$ or at $\infty$
either do not exists at all, or do not converge at the location of the
insertions.

\subsection{Basic commutative diagram}

In this situation, we may obtain a uniformizing map $f_{A\cup B}$ by
first removing $B$ by $f_B$, which is regular around $\infty$ and such
that $f_B(z)=z+O(1)$ at infinity, then $\tild{A} \equiv f_B(A)$ by
$f_{\tild{A}}$ which is regular around $0$ and fixes $0$ (as
mentionned before, $f_{\tild{A}}'(0)\neq 1$ is allowed).  Suppose that
$B$ is included in an open ball of radius $r$ and $\tild{A}$ is
included in the complement of a closed ball of radius $R$, both
centered at the origin. Now choose $z$ such that $|z| > r$ but
$|f_B(z)|<R$ \footnote{Such $z$'s exist in the above geometry, for
  instance in a small neighborhood of the segment of the real axis
  that separates $A$ and $B$. In such a region, radial ordering is
  also preserved by the maps.}. For such $z$'s, first the composition
$f_{A\cup B}(z)=f_{\tild{A}}\circ f_B(z)$ can be computed by inserting
the series expansions, and second
$G^{-1}_{f_{\tild{A}}}\left(G^{-1}_{f_B}T(z)G_{f_B}\right)G_{f_{\tild{A}}}$
is well defined, given by absolutely convergent series, and is equal
to $T(f_{A\cup B}(z))f'_{A\cup B}(z)^2+\frac{c}{12}Sf_{A\cup B}(z)$.

Of course, the roles of $A$ and $B$ could be interchanged, and we
could first remove $A$ by $f_A$ which is regular around $0$ and fixes
$0$ and then $\tild{B} \equiv f_A(B)$ by $f_{\tild{B}}$ which is
regular around $\infty$ and such that $f_{\tild{B}}(z)=z+O(1)$.

As they uniformize the same domain, we know that $f_{\tild{A}}\circ
f_B$ and $f_{\tild{B}}\circ f_A$ differ by a (real) linear fractionnal
transformation : there is an $h\in PSL_2(\mathbb{R})$ such that $
f_{\tild{B}}\circ f_A=h \circ f_{\tild{A}}\circ f_B$. Suppose that
$f_A$ and $f_B$ are given. There is some freedom in the choice of
$f_{\tild{A}}$ and $f_{\tild{B}}$ : namely we can replace
$f_{\tild{A}}$ by $h_0 \circ f_{\tild{A}}$ where $h_0$ is a linear
fractionnal transformation fixing $0$ , and $f_{\tild{B}}$ by
$h_{\infty} \circ f_{\tild{B}}$ where $h_{\infty}$ is a linear
fractionnal transformation such that $h_{\infty}(z)=z+O(1)$ at
infinity, i.e. a translation. A simple computation shows that unless
there is a $z$ such that $f_A(z)=\infty$ and $f_B(z)=0$, there is a
unique choice of $f_{\tild{A}}$ and $f_{\tild{B}}$ such that $
f_{\tild{B}}\circ f_A=f_{\tild{A}}\circ f_B$. In the sequel, we shall
concentrate on this generic situation. 
\begin{figure}[htbp]
  \begin{center}
    \includegraphics[width=1.\textwidth]{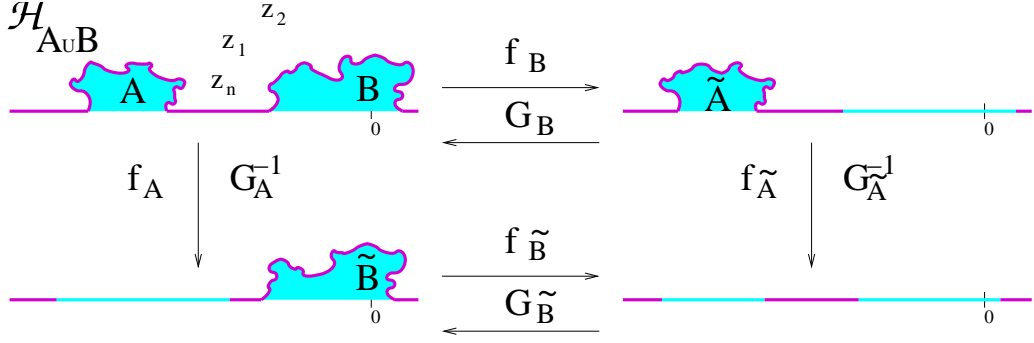}
      \caption{\em The generic commutative diagramm.}
      \label{fig:sleblobab}
  \end{center}
\end{figure}
So we deduce that for
$z$'s in some open set,
$G^{-1}_{f_{\tild{A}}}\left(G^{-1}_{f_B}T(z)G_{f_B}\right)G_{f_{\tild{A}}}=
G^{-1}_{f_{\tild{B}}}\left(G^{-1}_{f_A}T(z)G_{f_A}\right)G_{f_{\tild{B}}}$.
As the modes $L_n$ of $T$ generate all states in a highest weight
representation, the operators $G_{f_B}G_{f_{\tild{A}}}$ and
$G_{f_A}G_{f_{\tild{B}}}$ have to be proportional : they differ at
most by a factor involving the central charge $c$. We write
$$G_{f_B}G_{f_{\tild{A}}}=Z(A,B)\; G_{f_A}G_{f_{\tild{B}}},$$
or
\begin{equation}
  \label{eq:wik}
  G_{f_A}^{-1}G_{f_B}=Z(A,B)\; G_{f_{\tild{B}}}G_{f_{\tild{A}}}^{-1}.
\end{equation}
As implicit
in the notation, $Z(A,B)$ depends only on $A$ and $B$ : a simple
computation shows that it is invariant if $f_A$ is replaced by $h_0
\circ f_A$ and $f_B$ by $h_\infty \circ f_B$.

Formula (\ref{eq:wik}) plays for
the Virasoro algebra the role that Wick's theorem plays for
collections of harmonic oscillators.  

We call $Z(A,B)$ a partition
function for the following reason : we can write
\begin{eqnarray*}
 \left<
  \Omega \left| G^{-1}_{f_{\tild{A}}}G^{-1}_{f_B}\left(\cdots
     \hat{T}(z)  \cdots \right)_r
    G_{f_B}G_{f_{\tild{A}}} \right| \Omega \right> 
 & = & \\ & & \hspace{-5cm}\frac{1}{Z(A,B)}\left<
  \Omega \left| G^{-1}_{f_{\tild{B}}}G^{-1}_{f_A}\left(\cdots
     \hat{T}(z)  \cdots \right)_r
    G_{f_B}G_{f_{\tild{A}}} \right| \Omega \right>
\end{eqnarray*}
But $ \left| \Omega \right>$ is annihilated by
$\mathfrak{b}_+$ and $\left< \Omega \right| $ is
annihilated by $\mathfrak{n}_-$ so
$$\statav{\cdots T(z) \cdots}_{H_{A\cup B}} = \frac{1}{Z(A,B)}\left<
  \Omega \left| G^{-1}_{f_A}\left(\cdots
     \hat{T}(z)  \cdots \right)_r
    G_{f_B} \right| \Omega \right>,$$
and 
$$Z(A,B)=\left<
  \Omega \left| G^{-1}_{f_A}G_{f_B} \right| \Omega \right>.$$

\subsection{Computation of the partition function}

The computation of $Z(A,B)$ goes along the following lines.  If one
changes $A$ by a small amount, the variation of $f_A$ can be written as
$\delta f_A=v_A( f_A)$. In order to keep the initial properties of $A$
and $B$, we impose that $v_A$ is a vector field holomorphic in
the full plane but for cuts along the real axis, satisfies the
Schwartz reflexion principle 
($v_A(\overline{z})=\overline{v_A(z)}$), and is such that the open
disk of convergence of its power series expansion at $0$ contains
$\tild{B}$. Similar considerations hold if $B$ is distorted slightly, 
we write $\delta f_B=v_B( f_B)$ and $v_B$ satisfies corresponding
conditions. Then we know that 
\begin{eqnarray*}
\delta ( G^{-1}_{f_A}G_{f_B}) & = & \oint_0 v_A(u)T(u) du
G^{-1}_{f_A}G_{f_B}-G^{-1}_{f_A}G_{f_B}\oint_{\infty} v_B(v)T(v)dv \\
& = & Z(A,B) \Big(\oint_0 v_A(u)T(u) du G_{f_{\tild{B}}}
G^{-1}_{f_{\tild{A}}} - G_{f_{\tild{B}}}G^{-1}_{f_{\tild{A}}}
\oint_{\infty} v_B(v)T(v)dv\Big)
\end{eqnarray*}
By hypothesis, we can deform the small contour around $0$ to a contour
in a region
where $v_A$ and $f_{\tild{B}}$ have a convergent expansion, and the
small contour around $\infty$ to a contour
in a region where $v_B$ and
$f_{\tild{A}}$ have a convergent expansion. Then we may conjugate,
with the result
\begin{eqnarray*}
\frac{\delta ( G^{-1}_{f_A}G_{f_B})}{Z(A,B)} & = & G_{f_{\tild{B}}}
(\oint  v_A(u)(T(f_{\tild{B}}(u))
f'_{\tild{B}}(u)^2 +\frac{c}{12}Sf_{\tild{B}}(u))du \\ & - & \oint
 v_B(v)(T(f_{\tild{A}}(v))
f'_{\tild{A}}(v)^2 +\frac{c}{12}Sf_{\tild{A}}(v))dv)G^{-1}_{f_{\tild{A}}}
\end{eqnarray*}
Taking the vacuum expectation value yields 
$$ \delta \log Z(A,B) = \frac{c}{12}\Big(\oint  v_A(u)Sf_{\tild{B}}(u)du -
\oint v_B(v)Sf_{\tild{A}}(v)dv\Big).$$

The explicit value of $\log Z(A,B)$ can be computed by means of
several formul\ae . 

The most symmetrical ones are obtained if $A$
and $B$ are both described by integrating infinitesimal deformations
of  $\monh$. Consider two families of hulls, $A_s$
and $B_t$ that interpolate between the trivial hull and $A$ or $B$
respectively. We arrange that $f_{A_{s}}$ and $f_{B_{t}}$ satisfy the
genericity condition, so that unique $f_{A_{s,t}}$ and $f_{B_{t,s}}$
exist, which satisfy $f_{B_{t,s}} \circ
f_{A_s}=f_{A_{s,t}} \circ f_{B_{t}}$. 

\begin{figure}[htbp]
  \begin{center}
    \includegraphics[width=.7\textwidth]{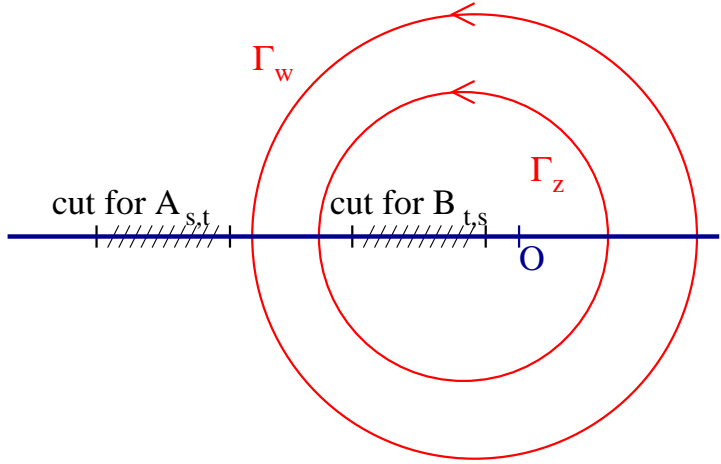}
      \caption{\em Integration contours intrication.}
      \label{fig:cut}
  \end{center}
\end{figure}

Define vector fields by $v_{A_{s}}$ and 
$v_{B_{t}}$ by $\frac{\partial f_{A_{s}}}{\partial
 s}=v_{A_{s}}(f_{A_{s}})$ and 
$\frac{\partial f_{B_{t}}}{\partial
 t}=v_{B_{t}}(f_{B_{t}})$. Now set $A_{s,t}=f_{B_t}(A_s)$ and
$B_{t,s}=f_{A_s}(B_t)$,
and define vector fields $v_{A_{s,t}}$ and 
$v_{B_{t,s}}$ by $\frac{\partial f_{A_{s,t}}}{\partial
 s}=v_{A_{s,t}}(f_{A_{s,t}})$ and 
$\frac{\partial f_{B_{t,s}}}{\partial
 t}=v_{B_{t,s}}(f_{B_{t,s}})$. Set
\begin{eqnarray}
\label{eq:anomsym}
L(A_{\sigma},B_{\tau}) &\equiv & \nonumber \\
& & \hspace{-2cm} \int_0 ^{\sigma} ds \int_0 ^{\tau}dt
\oint_{\Gamma_w} dw \oint_{\Gamma_z} dz \;
v_{A_{s,t}}(w)\frac{6}{(z-w)^4}v_{B_{t,s}}(z)
\end{eqnarray}
where the contours $\Gamma_w$ and $\Gamma_z$ are
simple contours in $\mathbb{C}$ of index $1$ with respect to $0$, such
that the bounded component of $\mathbb{C}\backslash \Gamma_z$ contains
the cuts of $f^{-1}_{B_{t,s}}$, the bounded component of
$\mathbb{C}\backslash \Gamma_w$ contains $\Gamma_z$ and the unbounded
component contains the cuts of $f^{-1}_{A_{s,t}}$ as described on
fig.(\ref{fig:cut}). We observe that the kernel is a four order pole,
i.e. is proportionnal to the
two-point correlation function for the stress energy tensor in the
plane geometry. We claim that
$$Z(A_{\sigma},B_{\tau}) = \exp {\frac{c}{12} L(A_{\sigma},B_{\tau})}.$$ 

This formula is very symmetrical, but it does not make clear that
$\log Z(A_{\sigma},B_{\tau})$ really depends only on $A_{\sigma}$ and
$B_{\tau}$, not on the full trajectories $A_s$, $s \leq \sigma$ and 
 $B_t$, $t \leq \tau$. The following steps are also useful to show
 that eq.(\ref{eq:anomsym}) has the correct variationnal derivative.
 
 We start by the change of variable $z=f_{A_{s,t}}(\zeta)$, which is valid
 for $z$ in a simply connected neighborhood of $\Gamma_w$ containing
 the origin, hence on $\Gamma_z$. Taking the $t$-derivative of
 $f_{B_{t,s}} \circ f_{A_s}=f_{A_{s,t}} \circ f_{B_{t}}$, we obtain
$$v_{B_{t,s}}(f_{A_{s,t}}(\zeta))=\frac{\partial f_{A_{s,t}}(\zeta)}{\partial
  t} + f_{A_{s,t}}'(\zeta)v_{B_{t}}(\zeta).$$
But $\frac{\partial f_{A_{s,t}}(\zeta)}{\partial t}$ is a holomorphic
function of $\zeta$ in a neighborhood of the origin containing the $\zeta$
integration contour, so in eq.(\ref{eq:anomsym}) we may replace
$v_{B_{t,s}}(z)dz$ by
$f_{A_{s,t}}'(\zeta)v_{B_{t}}(\zeta)f_{A_{s,t}}'(\zeta)d\zeta$. Hence
\begin{eqnarray}
\label{eq:anomzeta}
L(A_{\sigma},B_{\tau}) &= &  \int_0 ^{\sigma} ds \int_0 ^{\tau}dt
\oint_{\Gamma_w} dw \oint_{f_{A_{s,t}}^{-1}(\Gamma_z)} d\zeta \;
v_{A_{s,t}}(w)\frac{6f_{A_{s,t}}'(\zeta)^{2}}
{(f_{A_{s,t}}(\zeta)-w)^4}v_{B_{t}}(\zeta) \nonumber \\
& = & -\int_0 ^{\sigma} ds \int_0 ^{\tau}dt
 \oint_{f_{A_{s,t}}^{-1}(\Gamma_z)} d\zeta \; v_{A_{s,t}}'''
(f_{A_{s,t}}(\zeta)) f_{A_{s,t}}'(\zeta)^{2}v_{B_{t}}(\zeta)
\end{eqnarray}
In the second line, the $w$ integral has been computed by the residue
formula. This is legitimate because, by
hypothesis, $v_{A_{s,t}}(w)$ is holomorphic in the bounded component
of $\mathbb{C}\backslash {\Gamma_w}$. 
 
We can now make use of a useful identity for the variations of the Schwartzian
derivative. From its definition one checks that $S(f+\varepsilon
v(f))(f)=\varepsilon v'''(f)+O(\varepsilon^2)$. Combined with the
cocycle property $S(f+\varepsilon v(f))(z)dz^2=S(f+\varepsilon
v(f))(f)df^2+S(f)(z)dz^2$ this yields 
$$\frac{d}{d\varepsilon}S(f+\varepsilon
v(f))(z)_{|\varepsilon=0}=v'''(f(z))f'(z)^2.$$
Finally
\begin{eqnarray}
\label{eq:anompart}
L(A_{\sigma},B_{\tau}) &= &  -\int_0 ^{\sigma} ds \int_0 ^{\tau}dt
 \oint_{f_{A_{s,t}}^{-1}(\Gamma_z)} d\zeta \;
 \frac{d}{ds}Sf_{A_{s,t}}(\zeta)v_{B_{t}}(\zeta) \nonumber \\
& = & - \int_0 ^{\tau}dt
 \oint_{f_{A_{s,t}}^{-1}(\Gamma_z)} d\zeta \;
 Sf_{A_{\sigma,t}}(\zeta)v_{B_{t}}(\zeta).
\end{eqnarray}
The roles of $A_{\sigma}$ and $B_{\tau}$ could be interchanged to
remove the $\Gamma_w$ and $t$ integrations, leading to
\begin{eqnarray*}
L(A_{\sigma},B_{\tau}) & = & \int_0 ^{\sigma} ds
\oint dw \; v_{A_{s}}(w) Sf_{B_{\tau,s}}(w) \\ & = & -
\int_0 ^{\tau} dt \oint dz \; v_{B_{t}}(z) Sf_{A_{\sigma,t}}(z).  
\end{eqnarray*}
Using these formul\ae , it is apparent that $L(A_{\sigma},B_{\tau})$
does not depend on the detailed way the hulls are built : only the
final hulls count. It is also clear that setting $A\equiv A_{\sigma}$,
$A \cup \delta A= A_{\sigma+d\sigma}$, $B\equiv B_{\tau}$, $B \cup
\delta B= B_{\tau+d\tau}$, the variation of
$\frac{c}{12}L(A_{\sigma},B_{\tau})$ is exactly the one of $\log
Z(A,B)$. So we have proved
\begin{eqnarray*}
\log Z(A_{\sigma},B_{\tau}) & = & \frac{c}{12}\int_0 ^{\sigma} ds
\oint dw \; v_{A_{s}}(w) Sf_{B_{\tau,s}}(w) \\ & = & -\frac{c}{12}
\int_0 ^{\tau} dt \oint dz \; v_{B_{t}}(z) Sf_{A_{\sigma,t}}(z).  
\end{eqnarray*}

\subsection{Two explicit computations}

Let  $a$ and $b$ be real positive numbers

\subsubsection{Example 1 : two slits}

We define the hull $B_b$ to be
the segment $]i0,ib]$ and $A_a$ the segment $[ia,i\infty[$ in
$\monh$. Assuming that $0 \leq b < a \leq \infty$ we compute $L(A_a,B_b)$.

We interpolate between the empty hull and $B_b$ (resp. $A_a$) by
$B_{\beta}$, $\beta \in ]0,b]$ (resp. $A_{\alpha}$, $\alpha \in
[a,\infty[$).  To uniformize $\monh \backslash B_{\beta}$ we take the
map $f_{B_{\beta}}(z)=(z^2+\beta^2)^{1/2}$ and for $\monh \backslash
A_{\alpha}$ the map $f_{A_{\alpha}}(z)=(z^{-2}+\alpha^{-2})^{-1/2}$.
Observe that $f_{B_{\beta}}$ maps $A_{\alpha}$ to $A_{\gamma}$ where
$\gamma=(\alpha^2-\beta^2)^{1/2}$ while $f_{A_{\alpha}}$ maps
$B_{\beta}$ to $B_{\delta}$, where
$\delta=\frac{\alpha\beta}{(\alpha^2-\beta^2)^{1/2}}$. One checks that
$f_{B_{\delta}} \circ f_{A_{\alpha}} = \frac{1}{1-b^2/a^2}
f_{A_{\gamma}} \circ f_{B_{\beta}}$, so we get a commutative diagram
by taking $f_{A_{\alpha,\beta}} = \frac{1}{1-b^2/a^2} f_{A_{\gamma}}$
and $f_{B_{\beta,\alpha}} = f_{B_{\delta}}$.

Now $Sf_{A_{\alpha,\beta}}(z) = Sf_{A_{\gamma}}(z) =
-\frac{3(z^2+2\gamma^2)}{2(z^2+\gamma^2)^2}$ so
$Sf_{A_{a,\beta}}(z)=-\frac{3(z^2+2(a^2-\beta^2))}{2(z^2+a^2-\beta^2)^2}$.
On the other hand
$\frac{d}{d\beta}f_{B_{\beta}}=\frac{\beta}{f_{B_{\beta}}}$ so
$v_{B_{\beta}}(z)=\frac{\beta}{z}$. To resume,
$$
L(A_a,B_b) = -\int_0^b d\beta \oint dz
v_{B_{\beta}}(z)Sf_{A_{a,\beta}}(z)= \int_0^b d\beta \oint dz
\frac{\beta}{z} \frac{3(z^2+2(a^2-\beta^2))} {2(z^2+a^2-\beta^2)^2}.$$

The relevant $z$-integral encircles the singularity at $0$ and no
other, so  $L(A_a,B_b)= 3 \int_0^b d\beta \frac{\beta} 
{a^2-\beta^2}.$ Finally
$$L(A_a,B_b)=-\frac{3}{2}\log (1-b^2/a^2).$$

\subsubsection{Example 2 : a slit and a half disc}

We keep the definitions above for $A_a$, $A_{\alpha}$, $\alpha \in
[a,\infty[$) and $f_{A_{\alpha}}$. But now $B_b$ is the intersection
of the disc of center $0$ and radius $b$ with $\monh$, and we
interpolate between the empty hull and $B_b$ we use the half discs
$B_{\beta}$, $\beta \in ]0,b]$. To uniformize $\monh \backslash
B_{\beta}$ we choose the map $f_{B_{\beta}}(z)=z+\beta ^2/z$. Observe
that $f_{B_{\beta}}$ maps $A_{\alpha}$ to $A_{\gamma}$ where now
$\gamma=(\alpha^2-\beta^2)/\alpha$. The Schwartzian derivative is
insensitive to the precise normalization of $f_{A_{\alpha,\beta}}$, so
we can compute it by using $f_{A_{\gamma}}$ : $Sf_{A_{a,\beta}}(z) =
-\frac{3(z^2+2(a^2-\beta^2)^2/a^2)} {2(z^2+(a^2-\beta^2)^2/a^2)^2}$. On the
other hand $\frac{d}{d\beta}f_{B_{\beta}} =
\frac{f_{B_{\beta}}-\sqrt{f_{B_{\beta}^2 -4\beta^2}}}{\beta}$
so $v_{B_{\beta}}(z)=\frac{z-\sqrt{z^2-4\beta^2}}{\beta}$, where the
square root is defined to ensure the appropriate properties of
$v_{B_{\beta}}$ : this vector field is holomorphic in $\monh$ with
negative imaginary part, real on the real axis away from
the cut and satisfies the Schwarz reflexion principle. Hence
$$L(A_a,B_b) =\int_0^b d\beta \oint dz
\frac{z-\sqrt{z^2-4\beta^2}}{\beta} \frac{3(z^2+2(a^2-\beta^2)^2/a^2)}
{2(z^2+(a^2-\beta^2)^2/a^2)^2}.$$
The relevant $z$-integral encircles the cut $[-2\beta,2\beta]$ and no
other singularity. We may compute it with the help of the residue formula,
because the integrand is meromorphic in the unbounded component of the
complement of the integration contour, regular at infinity but with
double poles at $z=\pm i (a^2-\beta^2)/a$. The index is $-1$ for both,
and the residue is the same as well. 
This leads to
$$L(A_a,B_b) =3\int_0^b \frac{d\beta}{\beta}
\frac{\beta^2(\beta^2+2a^2)}{a^4-\beta^4}.$$
Finally
$$L(A_a,B_b)=\frac{3}{4}\log \frac{1+b^2/a^2}{(1-b^2/a^2)^3}.$$

We observe in these two examples that $L(A,B)$ becomes singular
when $A$ and $B$ have a contact. We also observe that $L(A,B)$ is
positive. There is a good reason for that. 

\subsection{Factorisation of unity and Virasoro vertex operators.}

Consider a hull $A$ whose closure does contain neither the origin
nor the infinity. There is a one parameter family of maps uniformizing
the complement of $A$ in $\mathbb{H}$ and which are regular both at
the origin and at infinity. Let us pick one of them, which we call
$f_A(z)$. Since $f_A(z)$ is regular at the origin, we may implement it
in conformal field theory by $G_{A^+}\,f_A'(0)^{-L_0}$ with 
$G_{A^+}$ in ${\cal N}_+$. Alternatively, since it is also regular at
infinity, we may implement it by $G_{A^-}\,f_A'(\infty)^{-L_0}$
with $G_{A^-}\in {\cal N}_-$. The product
$$
{\cal V}_A\equiv 
G_{A^-}\,f_A'(\infty)^{-L_0}\,f_A'(0)^{L_0}\,G_{A^+}^{-1}
$$
is the Virasoro analogue of what vertex operators of dual or string
models are for the Heisenberg or the affine Kac-Moody algebras. It
does not depend on the representative one chooses in the one parameter
family. This product is well defined and non trivial in positive
energy representation. It may be thought of as the factorization of
the identity since the conformal transformation it implements is the
composition of two inverse conformal maps.

\section{The partition function martingale}

We now come to the application that has motivated most of our
investment in the explicit implementation of conformal
transformations. For the convenience of the reader, we start with a
quick reminder of \cite{bb1} phrased in a more rigorous setting.
Remember that
$c_{\kappa}=\frac{(6-\kappa)(8\kappa-3)}{2\kappa}$ and
$h_{\kappa}=\frac{6-\kappa}{2\kappa}$. The Verma module
$V(c_{\kappa},h_{\kappa})$ is not irreducible, and
$(-2L_{-2}+\frac{\kappa}{2}L_{-1}^2)$ acting on the highest weight state
is another highest weight generating a subrepresentation. We quotient
$V(c_{\kappa},h_{\kappa})$ by this subrepresentation and denote by
$\ket{\omega}$ the highest weight state in the quotient. 
Then $(-2L_{-2}+\frac{\kappa}{2}L_{-1}^2)\ket{\omega}=0$. 

\subsection{Ito's formula for $G_{\monk _t}$}

The maps $\mony _t$ and $\monk _t=\mony _t-\xi _t$ that uniformize the
growing hull $\monK _t$ fix the point at infinity, so that there are
well defined elements $ G_{\mony _t}, G_{\monk _t} \in
\mathcal{N}_-\subset \overline{\mathcal{U} (\mathfrak{n}_{-})}$
implementing them in CFT. The maps are related by a change of the
constant coefficient in the expansion around $\infty$, so the operators
are related by $G_{\monk _t}=
G_{\mony _t}e^{\xi _t L_{-1}}$. The map $\mony _t$ satisfies the
ordinary differential equation $\partial_t
\mony_t(z)=\frac{2}{\mony_t(z)-\xi_t}$, the corresponding vector field
being $v(\mony)=\frac{2}{\mony-\xi_t}$ whose expansion at infinity
reads $v(\mony)=2\sum_{m\leq -2} f^{m+1}\xi_t^{-m-2}$, so that
$$G_{\mony _t}^{-1}dG_{\mony _t}=-2dt\sum_{m \leq -2} L_m
\xi_t^{-m-2}=-2e^{\xi _t L_{-1}}L_{-2}e^{-\xi _t L_{-1}}dt.$$
To get $G_{\monk _t}^{-1}dG_{\monk _t}$ it remains only
to compute the Ito derivative of $e^{\xi _t L_{-1}}$ which reads
$e^{-\xi _t L_{-1}}de^{\xi _t L_{-1}}=L_{-1}d\xi
_t+\frac{\kappa}{2}L_{-1}^2dt$.
Finally, 
$$G_{\monk _t}^{-1}dG_{\monk _t}=(-2L_{-2}+\frac{\kappa}{2}
L_{-1}^2)dt+ L_{-1}d\xi_t$$
 as announced in section \ref{sec:concft}.

In particular, $dG_{\monk _t}\ket{\omega}=L_{-1}d\xi_t G_{\monk
  _t}\ket{\omega}$, so that $G_{\monk _t}\ket{\omega}$ is a
(generating function of) local
martingale(s).

\subsection{The partition function martingale}

We have given an explicit formula for $Z(A,B)$, but motivated by the
martingale generating function, we shall sandwich
$G^{-1}_{f_A}G_{f_B}$ not with the vacuum
$\Omega$ but with another highest weight state, namely $\left |\omega
\right>$. Using the Virasoro Wick theorem, one computes that (remember
that $\bra{\omega}$ is 
  annihilated by $\mathfrak{n}_-$, but $\ket{\omega}$ is not
  annihilated by $\mathfrak{b}_+$, the $L_0$ part contributes)
$$\aver{G^{-1}_{f_A}G_{f_B}}=Z(A,B)\aver{G^{-1}_{f_{\tild{A}}}}=Z(A,B)
f_{\tild{A}}'(0)^{h_{\kappa}}.$$
Observe that while the vacuum expectation value depends only on the
hulls, the expectation value in a non conformally invariant state
depends on the choices of $f_A$ and $f_B$.

We apply the results of section \ref{sec:wick} to the case when $B$ is
the growing hull $K_t$ and $A$ is another disjoint hull. From the
previous computation we know that $\aver{G_{f_A}^{-1}G_{\monk _t}}$
is a local martingale.

We start from $f_A$ and $\mony _t$ to build a commutative diagram as
before, with maps denoted by $f_{\tild{A}_t}$ and $\tild{\mony} _t$
uniformizing respectively $\mony _t(A)$ and $f_A(K_t)$, and satisfying
$\tild{\mony} _t \circ f_A=f_{\tild{A}_t} \circ \mony _t$.
Now 
\begin{eqnarray*}
\aver{G_{f_A}^{-1}G_{\monk _t}} & = & \aver{G_{f_A}^{-1}G_{\mony
    _t}e^{\xi _t L_{-1}}} \\ & = &
  Z(A,K_t)\aver{G_{f_{\tild{A}_t}}^{-1}e^{\xi _t L_{-1}}}\\
& = & Z(A,K_t) \aver{\left(e^{-\xi _t
      L_{-1}}G_{f_{\tild{A}_t}}e^{f_{\tild{A}_t}(\xi _t) 
      L_{-1}}\right)^{-1}}.  
\end{eqnarray*}

From eq.(\ref{eq:trans}) we know that the operator $e^{-\xi _t
  L_{-1}}G_{f_{\tild{A}_t}}e^{f_{\tild{A}_t}(\xi _t) L_{-1}}$
corresponds to the map $z \mapsto
f_{\tild{A}_t}(\xi_t+z)-f_{\tild{A}_t}(\xi_t)$, so that
  $$\aver{\left(e^{-\xi _t
        L_{-1}}G_{f_{\tild{A}_t}}e^{f_{\tild{A}_t}(\xi _t) L_{-1}}\right)^{-1}}=f_{\tild{A}_t}'(\xi_t)^{h_{\kappa}}.$$

 From the Loewner equation
$v_{K_t}(z)=\frac{2}{z-\xi_t}$ and     
\begin{eqnarray*}L(A,K_t) & = & -
\int_0 ^{t} d\tau \oint dz \; \frac{2}{z-\xi_{\tau}}
  Sf_{A_{\tau}}(z)\\
& = & -2\int_0 ^{t} d\tau Sf_{A_{\tau}}(\xi_{\tau})
\end{eqnarray*}
Finally
$$\aver{G_{f_A}^{-1}G_{\monk
    _t}}=f_{\tild{A}_t}'(\xi_t)^{h_{\kappa}} \exp {-\frac{c}{6}\int_0
  ^{t} d\tau Sf_{A_{\tau}}(\xi_{\tau})},$$ 
were $f_{A_{\tau}}\circ \mony_{\tau}$ uniformizes the two hull geometry
corresponding to $A \cup K_{\tau}$ and  $f_{A_{\tau}}$ is normalized
to ensure the commutativity of the uniformization diagram as explained
before. It should be noted that the randomness in the above formula is
explicit through the appearance of $\xi_t$ but also implicit through
$f_{\tild{A}_{\tau}}$ which is a random function. 

This local martingale was discovered without any recourse to
representation theory by
Lawler, Schramm and Werner \cite{LSW}, but we hope to have convinced
the reader that it is nevertheless deeply rooted in CFT.

For the sake of completennes, we shall give two illustration of how
this machinery is used to compute explicit probabilities. The
following discussion does not claim originality, as the derivations
merely sketch the ones given in \cite{LSW}.   

\subsection{Restriction}

We already know that $$\aver{G_{f_A}^{-1}G_{\monk _t}}$$
is a local
martingale. One can show that it is a true martingale for $\kappa \leq
4$, let us just note that the region $\kappa \leq 4$ is also the one
for which, almost surely, the SLE hull $K_t$ is a simple curve that
avoids the real axis at all positive times. For the rest of this
section assume $\kappa \leq 4$. 

Suppose that $A$ is bounded and choose a very large semi circle
${\mathcal C }_R$ of radius $R$ in $\mathbb{H}$ centered at the
origin. Let $\tau_R$ be the first time when $K_t$ touches either $A$
or ${\mathcal C }_R$.  Then $\tau_R$ is a stopping time. It is crucial
to normalize $f_A$ correctly, and one does so by imposing that it
fixes $0$ (as already done) and that moreover $f_A(z)=z+O(1)$ close to
$\infty$, which by use of the commutative diagram ensures that ensures
that $f_{\tild{A}_t}(z)=z+O(1)$ close to $\infty$ as well. These three
conditions fix $f_A$ completely. Then we claim that
$f_{\tild{A}_{\tau_R}}'(\xi_{\tau_R})$ is $0$ if the SLE hull hits $A$
at $\tau_R$ and goes to $1$ for large $R$ if the SLE hull hits
${\mathcal C }_R$ at $\tau_R$. Indeed, when the hull approaches $A$,
one or more points on $\tild{A}_{t}$ approach $\xi_t$, and at the
hitting time, a bounded connected component is swallowed $\xi_t$ (this
uses the normalization of $f_A$) indicating that the derivative has to
vanish there. On the other hand, if ${\mathcal C }_R$ is hit first,
then $\tild{A}_{\tau_R}$ is dwarfed so that (this uses again the
normalization of $f_A$) $f_{\tild{A}_{\tau_R}}$ is close to the
identity map away from $\tild{A}_{\tau_R}$ and in particular at the
point $\xi_{\tau_R}$.  The behaviour of the other factor in the
martingale, $Z(A,K_t)$, is much harder to control, so we now restrict
to $\kappa=8/3$, which is the same as $c_{\kappa}=0$ because $\kappa
\leq 4$. So the partition function martingale
$f_{\tild{A}_t}'(\xi_t)^{h_{8/3}}$, at $t=\tau_R$ is $0$ if $A$ is hit
before ${\mathcal C }_R$ and close to $1$ if the opposite is true. But
the expectation of a martingale is time independant, so that the
probability that $K_t$ hits $A$ is
$f_{\tild{A}_t}'(\xi_t)^{h_{8/3}}_{|t=0}=f'_A(0)^{5/8}$.

\subsection{Locality}

Let us consider again the case when $B$ is the SLE hull $K_t$ and  $A$
another disjoint hull. We may apply the Virasoro Wick theorem to
$G^{-1}_{f_A}\, G_{\monk_t}$ to get
$$
G^{-1}_{f_A}\, G_{\monk_t}= 
Z(A,K_t)\ G_{\tilde \monk_t}\, G^{-1}_{f_{\widehat A_t}}
$$
Here $\tilde \monk_t$ is a uniformizing map of the image of the SLE
hull by $f_A$ and it defines the SLE growth in
$\mathbb{H}\setminus A$. Its lift $G_{\tilde \monk_t}$ in ${\cal
  N}_-$ depending locally on $\monk_t$ is random. A simple computation
shows that its Ito derivative is
$$
G_{\tilde \monk_t}^{-1} dG_{\tilde \monk_t}
=(-2L_{-2}+\frac{\kappa}{2}L^2_{-1}) f'_{\widehat A_t}(0)^2 dt
+ \frac{\kappa-6}{2}L_{-1}f''_{\widehat A_t}(0) dt+
L_{-1}f'_{\widehat A_t}(0) d\xi_t
$$
Hence, for $\kappa=6$, $G_{\tilde \monk_t}$ is statistically
equivalent to $G_{\monk_t}$ up to a time reparametrisation,
$dt\to ds = f'_{\widehat A_t}(0)^2 dt$. This expresses the locality
property of critical percolation.

\appendix

\section{Proof of identity (\ref{eq:zerocurv})}
\label{app:A}

We start with the proof of eq.(\ref{eq:zerocurv}) :
the operators $A_m \equiv  \sum_{n\geq m}
L_n \oint_0 dw  w^{m+1}\frac{f'(w)}{f(w)^{n+2}}$ satisfy the zero
curvature equation
$$\frac{\partial A_l}{\partial f_k}-\frac{\partial A_k}{\partial
  f_l}=[A_k,A_l].$$
Integration by parts gives
$$\frac{\partial}{\partial f_l}\oint_0 dw
w^{m+1}\frac{f'(w)}{f(w)^{n+2}}= 
\frac{m+1}{n+1}\frac{\partial}{\partial f_l}\oint_0 dw 
w^m\frac{1}{f(w)^{n+1}}=-(m+1)\oint_0 dw 
\frac{w^{l+m+1}}{f(w)^{n+2}}$$ so 
$$\frac{\partial A_l}{\partial f_k}-\frac{\partial A_k}{\partial
  f_l}=(k-l)\sum_j L_j \oint_0 dw
\frac{w^{k+l+1}}{f(w)^{j+2}}.$$
On the other hand, 
$$[A_k,A_l]=\sum_{m,n} (m-n)L_{m+n} \oint_0 du
u^{k+1}\frac{f'(u)}{f(u)^{m+2}} \oint_0 dv
v^{l+1}\frac{f'(v)}{f(v)^{n+2}}.$$
Split this sum in two pieces by splitting $m-n=(m+1)-(n+1)$. In the sum
involving $m+1$ use 
$$(m+1)\oint_0 du u^{k+1}\frac{f'(u)}{f(u)^{m+2}}=(k+1)\oint_0 du
\frac{u^k}{f(u)^{m+1}}.$$ In the sum involving $n+1$ use 
$$(n+1)\oint_0 dv v^{l+1}\frac{f'(v)}{f(v)^{n+2}}=(l+1)\oint_0 dv
\frac{v^l}{f(v)^{n+1}},$$ interchange the dummy variables $m$ and
$n$, and also $u$ and $v$. This leads to
$$[A_k,A_l]=\sum_{m,n} L_{m+n}\oint_0 du  \oint_0 dv
\frac{f'(v)((k+1)u^kv^{l+1}-(l+1)u^lv^{k+1})}{f(u)^{m+1}f(v)^{n+2}}.$$
Up to now, the contours in the $u$ and $v$ planes where
independent. But if they are adjusted in such a way that
$|f(v)|<|f(u)|$, we can fix $j=m+n$ and sum over $m$ to obtain
$$[A_k,A_l]=\sum_{j}L_j\oint_0 du  \oint_0 dv
\frac{f'(v)((k+1)u^kv^{l+1}-(l+1)u^lv^{k+1})}{(f(u)-f(v))f(v)^{j+2}}.$$
Inside the $u$-plane contour, the singularities of the $u$-integrand
consist now in a simple pole at
$u=v$, and taking the residue leads to
$$[A_k,A_l]=(k-l)\sum_j L_j \oint_0 dw
\frac{w^{k+l+1}}{f(w)^{j+2}}=\frac{\partial A_l}{\partial f_k}-
\frac{\partial A_k}{\partial f_l}.$$
This concludes the proof.

\section{Proof of identity (\ref{eq:conjl})}
 \label{app:B}

We continue with the proof of eq.(\ref{eq:conjl}) : 
$$G_f^{-1}L_mG_f=\frac{c}{12} \oint_0 dw w^{m+1}Sf(w)+\sum_{n\geq m}
L_n \oint_0 dw w^{m+1} \frac{f'(w)^2}{f(w)^{n+2}} \qquad m \in
\mathbb{Z}.$$

Observe that if we extend the summation over all $n$'s, the integrals
with $n < m$ vanish anyway. 
Defining $L_m(f)$ to be the right-hand side, one way to prove this
identity could be the tedious check that both sides have
the same variation when $f$ is changed into $f +\varepsilon z^{k+1}$, i.e.
$$\frac{\partial L_m(f)}{\partial f_k}= \left[\sum_{l\geq k} L_l
  \oint_0 dw w^{k+1}\frac{f'(w)}{f(w)^{l+2}} , L_m(f)\right] \qquad k
\geq 1.$$
This can be done, but it is simpler to consider the
variation of $G_f$ and $L_m(f)$ when $f$ is changed to $f +\varepsilon
v(f)$.  If $v(f)=\sum_{l\geq 1} v_l f^{l+1}$, we know that the
variation of $G_f$ is $-G_f\sum_{l\geq 1} v_lL_l$.  Now
\begin{eqnarray*}
[\sum_{l\geq 1} v_lL_l,\sum_{n}L_n \oint_0 dw w^{m+1}
\frac{f'(w)^2}{f(w)^{n+2}}] & = & \sum_{l\geq 1} v_l \sum_n
((l-n)L_{l+n} \\ & & \hspace{-1cm} +\frac{c}{12}
\delta_{l+n,0}(l^3-l))\oint_0 dw w^{m+1} \frac{f'(w)^2}{f(w)^{n+2}}]  
\end{eqnarray*}

For the term involving the central
charges we sum over $n$, then $l$ and get $\frac{c}{12}\oint_0 dw
w^{m+1} f'(w)^2 v'''(f(w))$.  For the remaining terms, for fixed $l$
we replace the dummy variable $n$ by $n-l$, leading to 
$$\sum_{l\geq 1}
v_l \sum_n (2l-n)L_n \oint_0 dw w^{m+1} \frac{f'(w)^2}{f(w)^{n-l+2}},$$
which is the same as 
$$\sum_n L_n \oint_0 dw w^{m+1} \frac{f'(w)^2
  (f(w)v'(f(w))-(n+2)v(f(w)))}{f(w)^{n+3}}.$$ 
Finally
\begin{eqnarray*}
\left[ \sum_{l\geq 1} v_lL_l, L_m(f) \right] & = & \sum_n L_n \oint_0 dw w^{m+1}
\frac{f'(w)^2 (f(w)v'(f(w))-(n+2)v(f(w)))}{f(w)^{n+3}} \\ & & + 
\frac{c}{12}\oint_0 dw w^{m+1}f'(w)^2 v'''(f(w)).   
\end{eqnarray*}
It is easily seen that this is nothing but $$\frac{dL_m(f+\varepsilon
  v(f))}{d\varepsilon}_{|\varepsilon=0},$$ which shows that
$$G_f^{-1}L_mG_f \quad \mathrm{and} \quad 
\frac{c}{12} \oint_0 dw w^{m+1}Sf(w)+\sum_{n\geq m}
L_n \oint_0 dw w^{m+1} \frac{f'(w)^2}{f(w)^{n+2}},$$ which coincide at
$f(z)=z$, have the same tangent map. Convexity ensures that they
coincide everywhere. 

\section{A few properties of $G_f$} 
\label{app:C}

The expansion of $G_f$ in powers of the $f_m$'s has an important
property that is already apparent in the expansion above. Let
$I=(i_1,i_2,\cdots)$ be a sequence of non negative integers with
finitely many nonzero terms. Let $E_m$ be the sequence made of zeroes
except for a single $1$ in the $m^{th}$ position, so that $I=\sum_m
i_mE_m$.  We define $|I|\equiv \sum_m i_m$ (which we call degree),
$d(I)\equiv \sum_m m i_m$ (which we call grading)
$I!\equiv\prod i_m! $, $f_{I}\equiv\prod_m f_m^{i_m}$ and
$L_I\equiv\prod_m L_m^{i_m}$ (with the convention that $L_1$ factors
are on the utmost right, then $L_2$, and so on). 

Then we claim that
$$G_f=\sum_I \frac{(-)^{|I|}}{I!}f_{I}(L_I+ \mathrm{lower \; order \;
  terms})$$
where ``lower order terms'' mean $f$-independent linear
combinations of $L_{J}$'s with $|J|<|I|$ but $d(J)=d(I)$. The same
statement would be true if we had chosen the opposite convention to
order the $L_m$'s in $L_I$. 

The statement that $d(J)=d(I)$ is simply that a dilation on $z$
multiplies $L_l$ by $\lambda ^l$ but divides $f_l$ by the same
factor. Alternatively, one can check that the factor $\oint_0 dw
w^{m+1}\frac{f'(w)}{f(w)^{n+2}}$ that appears in eq.(\ref{eq:pdesys}) is a
polynomial in the $f_l$'s of grading $n-m$. 

The proof $|J|<|I|$ that obtained by taking a commuting limit : we set
$f_m\equiv \varepsilon \varphi_m$ and $\Lambda_m \equiv \varepsilon
L_m$ (think of $\varepsilon$ as $\hbar$). Then in the limit
$\varepsilon \rightarrow 0$ keeping the $\varphi_m$'s fixed, on the
one hand the $\Lambda_m$'s commute , and on the other hand $\oint_0 dw
w^{m+1}\frac{f'(w)}{f(w)^{n+2}}=\delta_{n,m}$ so that the differential
system defining $G_f$ reduces to $\frac{\partial G_f}{\partial
  \varphi_m}=-G\Lambda_m$, with solution $G_f=e^{-\sum_m \varphi_m
  \Lambda_m}$. This implies that in the $\varepsilon$ expansion in
terms of $\varphi_m$'s and $\Lambda_m$'s, $G_f=\sum_I
\frac{(-)^{|I|}}{I!}\varphi_{I}(\Lambda_I+ O(\varepsilon))$. But
expressed in terms of $f_m$'s and $L_m$'s the result is
$\varepsilon$-independent. This means that the coefficient of
$\varphi_{I}\varepsilon ^k$ involves only $\Lambda_J$'s with
$|J|=|I|-k$. This concludes the proof. 

An analogous computation would show that
$$G_f^{-1}=\sum_I \frac{1}{I!}f_{I}(L_I+ \mathrm{lower \; order \;
  terms}).$$

We can rephrase these results as follows :  
\begin{eqnarray}
  \label{eq:polf}
G_f & = & \sum_I \frac{(-)^{|I|}}{I!}L_I(f_{I}+ \mathrm{higher \; order \;
  terms}), \\ 
G_f^{-1} & = & \sum_I \frac{1}{I!}L_I(f_{I}+ \mathrm{higher \; order \;
  terms}),\end{eqnarray}
where ``higher order terms'' mean $L$-independent linear
combinations of $f_{J}$'s with $|J|>|I|$ but $d(J)=d(I)$. 
In particular the polynomials in the $f_m$'s that appear as
coefficients of the $L_I$'s in the above expansions form a basis of the
space of all polynomials in the $f_m$'s. 

These observations will be useful
for the application to representation theory in section \ref{sec:rep}. 

We can also write down a general recursive formula.  We define
$P_I$ by $G_f\equiv\sum_I \frac{(-)^{|I|}}{I!}f_{I}P_I$ and combinatorial
coefficients $C_J(m,n)$ by $\oint_0 dw
w^{m+1}\frac{f'(w)}{f(w)^{n+2}}\equiv \sum_J
\frac{(-)^{|J|}}{J!}f_{J}C_J(m,n)$. The integrand can we written as
$w^{m-n}\frac{dw}{w}$ times a function in which each $f_l$ is
multiplied by $z^l$ : $C_J(m,n)=0$ unless $d(J)=n-m$.
The partial differential equations for $G_f$ lead to
difference equations for the $P_I$'s. One gets
$$P_{K+E_m}=\sum_{I+J=K} \frac{K!}{I!J!} C_J(m,m+d(J))P_IL_{m+d(J)}.$$
One finds $P_{E_m}=L_m$, $P_{E_m+E_n}=L_mL_n+(n+1)L_{m+n}$, $\cdots$.

\section{Final steps for the proof of (\ref{eq:diffn<1f})}
\label{app:D}

We start from eq.(\ref{eq:diffn<1fd}), repeated here for convenience :
\begin{eqnarray*}
P_{L_ny}+\left(\frac{c}{12} \oint_0 dw
f(w)^{n+1}\frac{Sf(w)}{f'(w)}- h\oint_0 dw
 \frac{f(w)^{n+1}}{w^{2}f'(w)}\right)P_{y} & = & \\ \hspace{-2cm}
-\sum_{m =n}^0 \oint_0 dw 
\frac{f(w)^{n+1}}{w^{m+2}f'(w)}\sum_{l \geq 1} \oint_0 dz z^{m+1}
\frac{f'(z)^2}{f(z)^{l+2}} \left<  G_f L_l y,x \right>.   
\end{eqnarray*}

Now, fix $m$ and concentrate on  $\sum_{l \geq 1} \oint_0 dz z^{m+1}
\frac{f'(z)^2}{f(z)^{l+2}} \left<  G_f L_l y,x \right>$. From the
Lagrange formula, one can expand  $z^{m+1}f'(z)$ in powers of $f(z)$
as
$$z^{m+1}f'(z)=\sum_{k \geq m} f(z)^{k+1}\oint_0 du
\frac{u^{m+1}f'(u)^2}{f(u)^{k+2}}.$$
Define 
\begin{equation} \label{eq:defhm}
 h_m(z)\equiv z^{m+1}f'(z)-\sum_{k, \; m \leq k \leq 0} f(z)^{k+1}\oint_0 du
\frac{u^{m+1}f'(u)^2}{f(u)^{k+2}}.\end{equation}
By definition, $h_m(z)$ is a $O(z^2)$ and its $z$ expansion reads
$$ h_m(z)=\sum_{j \geq 1} z^{j+1} \left( f_{j-m}(j-m+1)-\sum_{k =
m}^{0}\oint_0 du \frac{u^{m+1}f'(u)^2}{f(u)^{k+2}}\oint_0 dv 
\frac{f(v)^{k+1}}{v^{j+2}}\right).$$
On the other hand, by construction, $h_m(z)$ is such that 
\begin{equation} \label{eq:prophm} \oint_0 dz
  z^{m+1}\frac{f'(z)^2}{f(z)^{l+2}}=\oint_0 dz 
\frac{h_m(z)f'(z)}{f(z)^{l+2}} \quad \mathrm{for} \quad l=1,2,\cdots.
\end{equation}
so, using again eq(\ref{eq:pdesys}),
\begin{eqnarray*}
\sum_{l \geq 1} \oint_0 dz 
z^{m+1} \frac{f'(z)^2}{f(z)^{l+2}} G_f L_l  & = & \\ & & \hspace{-4.5cm}
-\sum_{j \geq 1} \left( f_{j-m}(j-m+1)-\sum_{k =
m}^{0}\oint_0 du \frac{u^{m+1}f'(u)^2}{f(u)^{k+2}}\oint_0 dv 
\frac{f(v)^{k+1}}{v^{j+2}}\right)\frac{\partial G_f}{\partial f_j}.
\end{eqnarray*}

\end{document}